\documentclass[11pt]{article}
\usepackage{amssymb, amsmath,graphicx,color}
\usepackage[top=3cm, bottom=3cm, left=3cm, right=3cm]{geometry}
\usepackage[square,numbers]{natbib}

\definecolor{webblue}{rgb}{0, 0, 0.5} 
\usepackage[pdfpagelabels,  
plainpages=true,
bookmarks=true,
bookmarksnumbered=true,
hypertexnames=true,
breaklinks=true,%
linkcolor=webblue,]{hyperref}
\hypersetup{
colorlinks  = true,
linkcolor = webblue,
urlcolor = webblue,
citecolor = webblue,
pdfauthor   = {},
pdfsubject = {},
pdftitle    = {},
pdfkeywords = {},
}

\renewcommand{\d}{\mathrm{d}}
\newcommand{\e}{\mathrm{e}}

\newcommand {\pdd}[2]{\frac{\partial #1}{\partial #2}}

\renewcommand{\v}{\mathbf{v}}

\newcommand{\n}{{\mathbf{n}}}
\renewcommand{\t}{{\mathbf{t}}}

\newcommand{\nablas}{\nabla_\perp}

\newcommand{\R}{\mathbf{R}}
\newcommand{\U}{\mathbf{U}}
\renewcommand{\u}{\mathbf{u}}
\renewcommand{\v}{\mathbf{v}}

\newcommand{\uuline}[1]{\underline{\underline{#1}}}
\newcommand{\tensor}[1]{\uuline{\boldsymbol{#1}}}
\renewcommand{\vector}[1]{{\mathbf{#1}}}
\newcommand{\stress}{\tensor{\sigma}}

\title{The viscoelastic rheology of transient diffusion creep}
\author{John F. Rudge\\
\small{Bullard Laboratories, Department of Earth Sciences, 
University  of 
Cambridge}}

\begin{document}

\maketitle

\begin{abstract}
Polycrystalline materials have a viscoelastic rheology where the strains 
produced by stresses depend on the timescale of deformation. Energy can be 
stored elastically within grain interiors and dissipated by a variety of 
different mechanisms. One such dissipation mechanism is 
diffusionally-accommodated/-assisted grain boundary sliding, also known as 
transient diffusion creep. Here we detail a simple reference model of transient 
diffusion creep, based on finite element calculations with simple grain 
shapes: a regular hexagon in 2D and a tetrakaidecadedron in 3D. The linear 
viscoelastic behaviour of the finite element models can be well described by a
parameterised extended Burgers model, which behaves as a Maxwell model at low 
frequencies and as an Andrade model at high frequencies. The parametrisation 
has a specific relaxation strength, Andrade exponent and Andrade time. The 
Andrade exponent depends only on the angles at which grains meet at triple 
junctions, and can be related to the exponents of stress singularities that 
occur at triple junctions in purely elastic models without diffusion. A 
comparison with laboratory experiments shows that the simple models here 
provide 
a lower bound on the observed attenuation. However, there are also clearly 
additional dissipative processes occurring in laboratory experiments that are 
not 
described by these simple models.
\end{abstract}

\section{Introduction}

A fundamental question of material science is to understand how a 
material deforms when it is stressed, i.e. what is the rheology of a 
material? There are two main approaches to improving our understanding of 
rheology: one is 
through careful laboratory experiments, the other is through building models 
of 
physical processes that occur within materials. It is only through 
modelling work, and through understanding the fundamental physics, that 
results from 
experiments can be safely extrapolated into parameter regimes beyond 
those 
of the laboratory. This is of particular concern in Earth sciences where the 
length 
scales and time scales of the laboratory are so small compared to those 
relevant for the planet 
and its long history.

The aim of the present manuscript is to provide a simple
grain-scale model of rock viscoelasticity that can be used as a basis for 
interpreting and extrapolating laboratory experiments. The model describes the 
effective viscoelastic behaviour which arises from the grain-scale diffusion of 
matter by physical mechanisms referred to as 
diffusionally-assisted or diffusionally-accommodated grain-boundary 
sliding \citep{Jackson2015,Takei2017,Cooper2002}; mechanisms also referred to 
as  
transient 
diffusion creep. There is a long history of models of transient diffusion creep, 
with early pioneering work by Lifshits, Raj and Ashby  
\citep{Lifshits1965,Raj1971,Raj1975}. The 
present work 
builds 
most closely on the more recent study by Lee et al. \citep{Lee2011}, who 
developed a 
two-dimensional finite element model of a bicrystal consisting of two elastic 
grains with a grain boundary between. The grain boundary allows 
for both sliding and diffusion of 
matter. Here, we model the same physics as Lee et al. \citep{Lee2011} but in a 
different 
geometry, considering an infinite tiling of hexagonal grains in two-dimensions 
and a tessellation of tetrakaidecahedral grains in three-dimensions, in turn 
building on models for steady-state diffusion creep described in 
\citep{Rudge2018a,Rudge2021}. A homogenisation approach is used to determine 
the effective 
upscaled viscoelastic tensors which describe the macroscale rheology arising 
from the grain-scale physics.  

The manuscript is organised as follows. The next section describes the model, 
and results for hexagonal grains are given in \autoref{sec:hex_res} and 
results for tetrakaidecahedral grains in \autoref{sec:tetra_res}. This is 
followed by comparison to laboratory data and previous studies in 
\autoref{sec:disc} and conclusions in \autoref{sec:conc}. A series of 
appendices provide the mathematical basis for the modelling work.

\section{Model description}

The model geometry consists of a tessellation of identical grains. The interior 
of grains 
are assumed to deform linear elastically, with intrinsic shear modulus $\mu$ 
and Poisson's ratio $\nu$. Grains are assumed to slide freely at the grain 
boundaries (zero shear stress). Vacancies and atoms can diffuse along the grain 
boundaries due to gradients in normal stress, leading to the plating out or 
removal of material along the grain boundaries. Energy is stored elastically 
within the grain interiors and dissipated by grain boundary diffusion. The 
medium thus behaves viscoelastically. The linear viscoelastic response of a 
material can be described in a number of different ways, but one of the most 
useful is as a function of frequency $\omega$. The strain response 
$\epsilon(t)$ of a viscoelastic material to a time harmonic shear stress 
$\sigma(t) = \sigma_0 \e^{i \omega t}$ is
\begin{equation}
 \epsilon(t) = J^* (\omega) \sigma(t)
\end{equation}
where $J^* (\omega)$ is the complex shear compliance. The reciprocal 
$G^*(\omega) = 1 / J^* (\omega)$ is known as the complex shear
modulus. As a complex quantity, $J^* 
(\omega)$ can be split into real and imaginary parts as $J^* 
(\omega) = J_1(\omega) - i J_2(\omega)$, where $J_1$ is termed the storage 
compliance and $J_2$ is termed the loss compliance. Finite element analysis 
allows the calculation of the function $J^* 
(\omega)$ for a given grain geometry, which is the main outcome of the 
present study. A detailed account of the governing equations and 
their discretisation by the finite element method can be found in appendices 
\ref{sec:govern_eq} to \ref{sec:num_imp}.

\section{Hexagonal grains}\label{sec:hex_res}

In general the viscoelastic response of a material is described by a 
fourth-rank complex compliance tensor, which in three-dimensions involves 
an array of 81 complex numbers as a function of frequency. However, the number 
of independent components is greatly reduced on consideration of symmetry, 
both in the governing equations and in the geometry. In 2D, a tiling of regular 
hexagons yields fourth rank tensors which are isotropic, i.e. there is no 
dependence on orientation, and the viscoelastic behaviour can be described by 
just two numbers as a function of frequency: a complex bulk compliance and a 
complex shear compliance. The assumed grain scale physics does not allow for 
any 
dissipation under bulk deformation and so the only viscoelastic behaviour that 
need be explicitly computed is that under shear.

Let us consider the end-member behaviour at zero and infinite frequency. At zero 
frequency (infinite period) the 
diffusion process reaches a steady-state. This steady-state deformation is 
steady-state 
diffusion creep (Coble creep \citep{Coble1963}) and the material behaves as a 
Newtonian viscous 
fluid. The steady-state shear viscosity $\eta$ for an array of hexagonal grains 
has been calculated by many authors \citep[e.g.][]{Spingarn1978, 
Cocks1990, Rudge2018a} and is
\begin{equation}
 \eta = \frac{1}{144} \frac{k T d^3}{\delta D_{gb} \Omega}
\end{equation}
where $k$ is Boltzmann's constant, $T$ is temperature, $d$ is the distance 
between opposite sides of the hexagon, $\delta$ is the grain boundary width, 
$D_{gb}$ is the self-diffusion coefficient for grain boundary diffusion, and 
$\Omega$ is the atomic volume. 

At the other extreme, at infinite frequency, diffusion can be neglected and the 
behaviour is that of an elastic solid. However, the effective shear 
modulus is less than the intrinsic shear modulus $\mu$ of the grains because we 
assume shear stresses are zero on the grain boundaries. We denote this 
effective modulus as $G_0$ and refer to it as the unrelaxed modulus, as there 
has been no relaxation of the 
stresses normal to the grain boundaries by the process of main interest 
here, by grain boundary diffusion.

\begin{figure}
\centering
 \includegraphics[width=0.59\columnwidth]{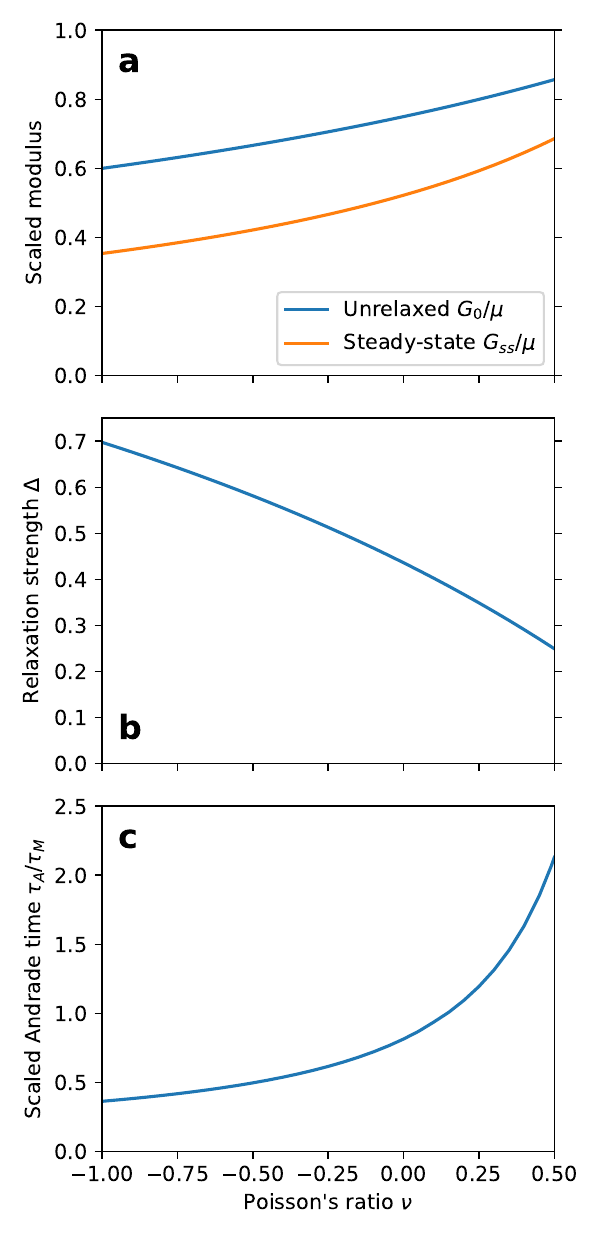}
 \caption{Properties of the model for hexagonal grains as a function of 
Poisson's ratio $\nu$. a) Scaled unrelaxed shear modulus $G_0/\mu$ and scaled
steady-state shear modulus $G_{ss}/\mu$. b) Relaxation strength $\Delta$ 
associated with grain boundary 
diffusion, $\Delta = (G_0/G_{ss}) - 1$. c) Ratio of the Andrade time $\tau_A$ to 
the Maxwell time $\tau_M$. The data in these plots can be well fit by simple 
rational functions of $\nu$ (see \autoref{sec:rational}). The plots show the 
behaviour for the full allowable range of Poisson's ratio for linear 
elasticity, $-1 \leq \nu \leq \tfrac{1}{2}$; a typical value for 
rock is $\nu \sim 0.3$.}
 \label{fig:hex_moduli}
\end{figure}

\autoref{fig:hex_moduli}a shows $G_0 / \mu$ as a function of Poisson's ratio 
$\nu$. This quantity was calculated previously for hexagonal grains by Gharemani
\citep{Ghahremani1980}; the 
results here are in complete agreement for 
the range of Poisson's ratios used in that study (see Figure 4 of 
\citep{Ghahremani1980}). $G_0 / \mu$ ranges from 0.6 to 0.86 depending on 
Poisson's ratio. For a typical value of Poisson's ratio of rock of around $\nu 
= 0.3$, the ratio $G_0 / \mu = 0.81$. 

The ratio of the steady-state viscosity $\eta$ to the unrelaxed modulus $G_0$ 
defines a 
timescale
\begin{equation}
 \tau_M = \frac{\eta}{G_0}
\end{equation}
which we refer to as the Maxwell time. By scaling the governing equations it 
can be shown that the Maxwell time is 
the only natural timescale in the problem, and hence it is useful to present 
all 
results in terms of this timescale, i.e. in terms of a dimensionless 
frequency $\omega \tau_M$. 

\begin{figure}
 \includegraphics[width=\columnwidth]{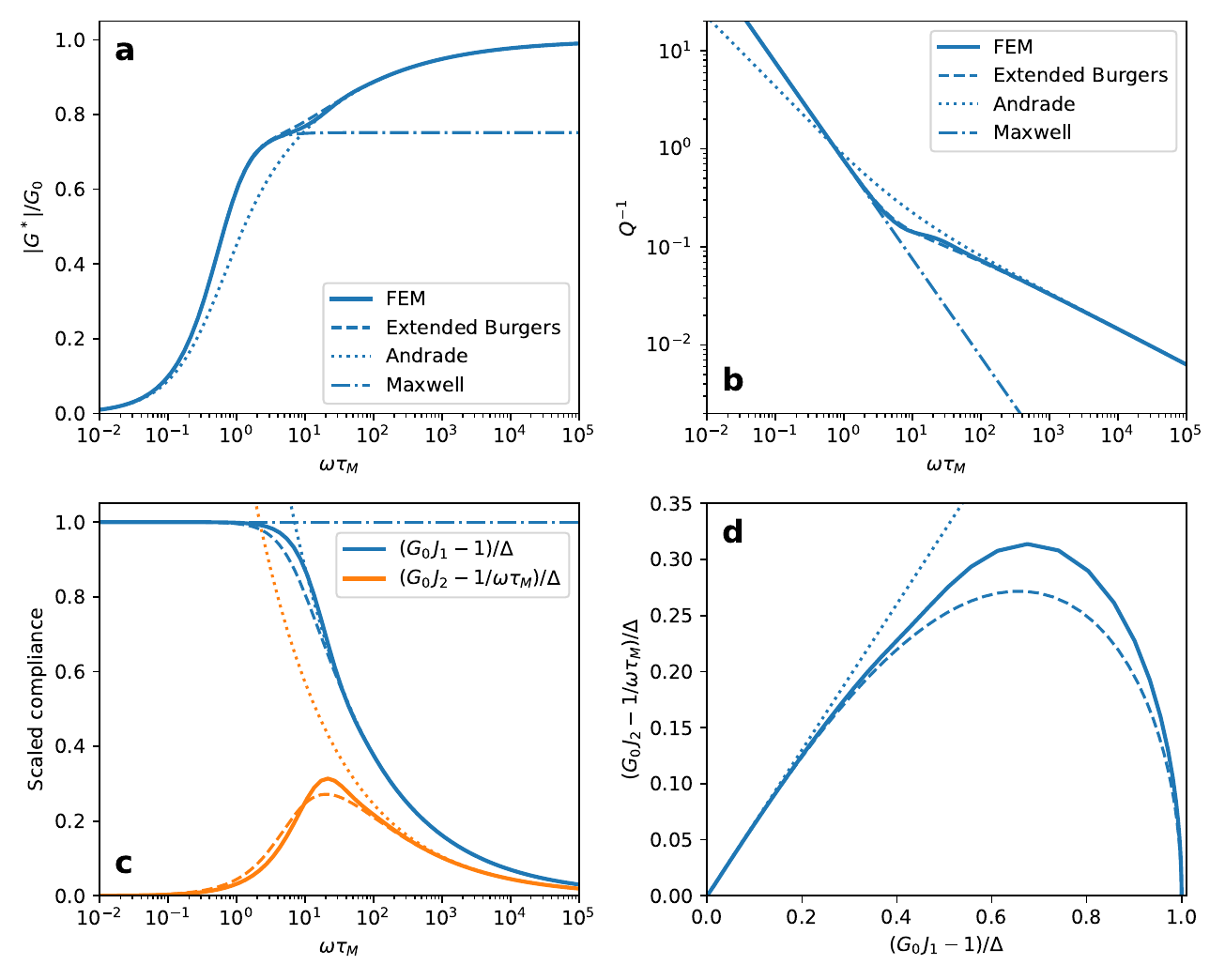}
 \caption{A summary of the viscoelastic response for hexagonal grains with a
Poisson's ratio $\nu = 0.3$. a) $|G^*|/G_0$, the ratio of the absolute shear 
modulus to the unrelaxed modulus as a function of dimensionless frequency 
$\omega \tau_M$. b) $Q^{-1}$, inverse quality factor (attenuation) as a 
function 
of dimensionless frequency. c) Scaled compliances $(G_0 J_1 - 1)/\Delta$ and 
$(G_0 J_2 
- 1/\omega \tau_M)/\Delta$ as a function of dimensionless frequency. d) 
Cole-Cole 
plot of the scaled compliances against each other. In all plots solid lines 
show the finite element solutions, dashed lines an extended Burgers model fit, 
dotted lines an Andrade model fit, dash-dotted lines a Maxwell model fit.}
 \label{fig:hex_all}
\end{figure}

\autoref{fig:hex_all} shows plots which summarise the viscoelastic behaviour 
of 
the model for hexagonal grains. The top two panels show the scaled
absolute shear modulus $|G^*|/G_0$ and the attenuation $Q^{-1} \equiv J_2/J_1$, 
quantities commonly 
plotted in experimental studies of forced oscillation; the bottom two panels 
show scaled versions of the compliances which illustrate the nature of the  
relaxation. \autoref{fig:hex_all} also shows fits to the finite element results 
using some simple viscoelastic models: the Maxwell model, the Andrade model 
and the extended Burgers Model (see 
\autoref{sec:simplemodels} for a description of these models). 

The behaviour of the model is best understood by considering its extremes. As 
remarked earlier, the behaviour at zero frequency is that of a viscous fluid. 
The 
behaviour at low frequencies $\omega \tau_M \ll 1$ more generally (referred 
to as the 
diffusionally-accommodated regime by \cite{Jackson2014,Jackson2015}) is that 
of a 
Maxwell model with a dashpot of viscosity $\eta$ in series with a spring of 
compliance $J_{ss}$ or modulus $G_{ss} = 1/J_{ss}$. $\eta$ is the steady-state 
viscosity and $G_{ss}$ is the steady-state modulus. The steady-state compliance 
 $J_{ss}$ is the limiting value of the storage 
compliance: $J_1 \rightarrow J_{ss}$ as $\omega \tau_M \rightarrow 0$. The 
relaxation strength 
$\Delta$ is the fractional increase in compliance due to the relaxation of 
stresses by grain boundary diffusion, given by
\begin{equation}
 \Delta = \frac{J_{ss}}{J_0} - 1 = \frac{G_0}{G_{ss}} - 1.
\end{equation}
The relaxation strength is plotted in \autoref{fig:hex_moduli}b and varies with 
Poisson's ratio $\nu$ from 
$\Delta=0.70$ to 0.25, with a value of $\Delta=0.33$ for a typical rock value of
$\nu=0.3$.

At high frequencies (the 
diffusionally-assisted regime of \cite{Jackson2014,Jackson2015}), with $\omega 
\tau_M \gg 1$, the behaviour is that 
of an Andrade 
model, with power-law variation of attenuation as $Q^{-1} \propto 
\left(\omega \tau_M \right)^{-\alpha}$ for a particular exponent $\alpha$. The 
Andrade exponent $\alpha$ can be calculated explicitly using singularity 
analysis 
\citep{Picu1996,Williams1952,Sinclair2004} following a scaling argument due to 
Lee et al.
\citep{Lee2011}. In the absence of grain boundary diffusion, the pure elastic 
problem with 
freely-sliding grain-boundaries has singularities in the stress at the triple 
junctions where three grains meet, i.e. the normal stresses go as $\sigma_{nn} 
\propto r^{-\lambda}$ where $r$ is the distance from the triple junction. The 
analysis of Picu and Gupta
\citep{Picu1996} shows that for a triple junction with equal angles of 120° and 
freely 
sliding grain boundaries the exponent $\lambda = 0.4491862$.

In the presence of diffusion, the singularity in the stress is removed. The 
length scale over which diffusion acts depends on the frequency of the 
forcing 
and scales with (\autoref{sec:lee_scaling})
\begin{equation}
 l = \left(\frac{\mu \delta D_{gb} \Omega}{k T \omega} \right)^{1/3}. 
\label{eq:ldef}
\end{equation}
Provided $l \ll d$ the dissipation due to grain boundary diffusion is 
concentrated in a small neighbourhood of the triple junction, whereas
elastic energy is stored throughout the grain interior. Scaling analysis shows 
that the power law exponent $\alpha$ in the Andrade model 
is related to the stress singularity exponent $\lambda$ as
\begin{equation}
 \alpha = \frac{2}{3} \left(1 - \lambda \right).
\end{equation}
For a derivation see \autoref{sec:lee_scaling} and section 4 of 
\citep{Lee2011}. From the Picu and Gupta result \citep{Picu1996}, the exponent 
$\alpha = 
0.3672092$.

The power law behaviour can be clearly seen in \autoref{fig:hex_all}, both in 
the high frequency end of the attenuation plot in panel b, but also in the 
scaled compliance plots in panels c and d. In panel c the power-law behaviour 
manifests as heavy tails at high frequencies, which can be contrasted with a 
simple Debye peak one would get for a Burgers rheology with a single 
relaxation time. Similarly, the behaviour near the origin in the Cole-Cole plot 
of panel d is tangent to a line with slope $\tan \pi \alpha/2$, quite different 
from a Burgers rheology which would plot as a semicircle on a Cole-Cole 
diagram.   

The other parameter which specifies the Andrade model is the Andrade time 
$\tau_A$ (appendix~\ref{sec:andrade}). Unlike the exponent $\alpha$, this 
time cannot be determined purely 
by examining the singular behaviour at the triple junctions, but is 
also sensitive 
to the overall grain geometry and Poisson's ratio. \autoref{fig:hex_moduli}c 
plots 
the Andrade time calculated from the finite element analysis as a function of 
Poisson's 
ratio, and shows that the Andrade time is broadly similar to the Maxwell time, 
increasing from $\tau_A/\tau_M=0.36$ to 2.13. The dimensionless ratio 
$\tau_A/\tau_M$ is sometimes referred to as the Andrade $\zeta$ parameter in 
tidal heating studies \citep{Bierson2024}. For a typical rock value of $\nu 
= 0.3$, $\tau_A/\tau_M = 1.315$. 

The Maxwell model and the Andrade model characterise the behaviour of the 
finite element 
results at low and high dimensionless frequency respectively. Also shown in 
\autoref{fig:hex_all} is another model fit to the finite element data using an 
extended 
Burgers Model (appendix \ref{sec:eb}, \cite{Jackson2010}). This model is 
similar to the Andrade 
model in 
having a relaxation spectrum with a power-law distribution of relaxation times, 
but has a long-period cut-off to the relaxation spectrum given by a parameter 
$\tau_H$ which can be related to $\tau_A$ and $\Delta$ by $\tau_H = \tau_A 
\left(\Gamma(1-\alpha) \Delta\right)^{1/\alpha}$, where $\Gamma(z)$ is the 
Gamma function. The extended Burgers model 
has the same asymptotic behaviour as the 
finite element model at both low and high frequencies. As can be seen in 
\autoref{fig:hex_all}, the extended Burgers model provides an excellent fit to 
the finite element results, with only subtle differences around $\omega \tau_M 
= 20$ to be 
seen in panels a and b which show absolute shear modulus and attenuation. The 
differences are somewhat more noticeable in the plots of scaled compliance in 
panels c 
and d, but even in these plots the differences are slight. Thus an effective 
parametrisation of the finite element results is to use an extended Burgers 
model with the 
appropriate values of $G_0$, $\tau_M$, $\Delta$ (\autoref{fig:hex_moduli}b), 
$\alpha=0.3672092$ and $\tau_A/\tau_M$ (\autoref{fig:hex_moduli}c).

\section{Tetrakaidecahedral grains}\label{sec:tetra_res}

\begin{figure}
 \includegraphics[width=\columnwidth]{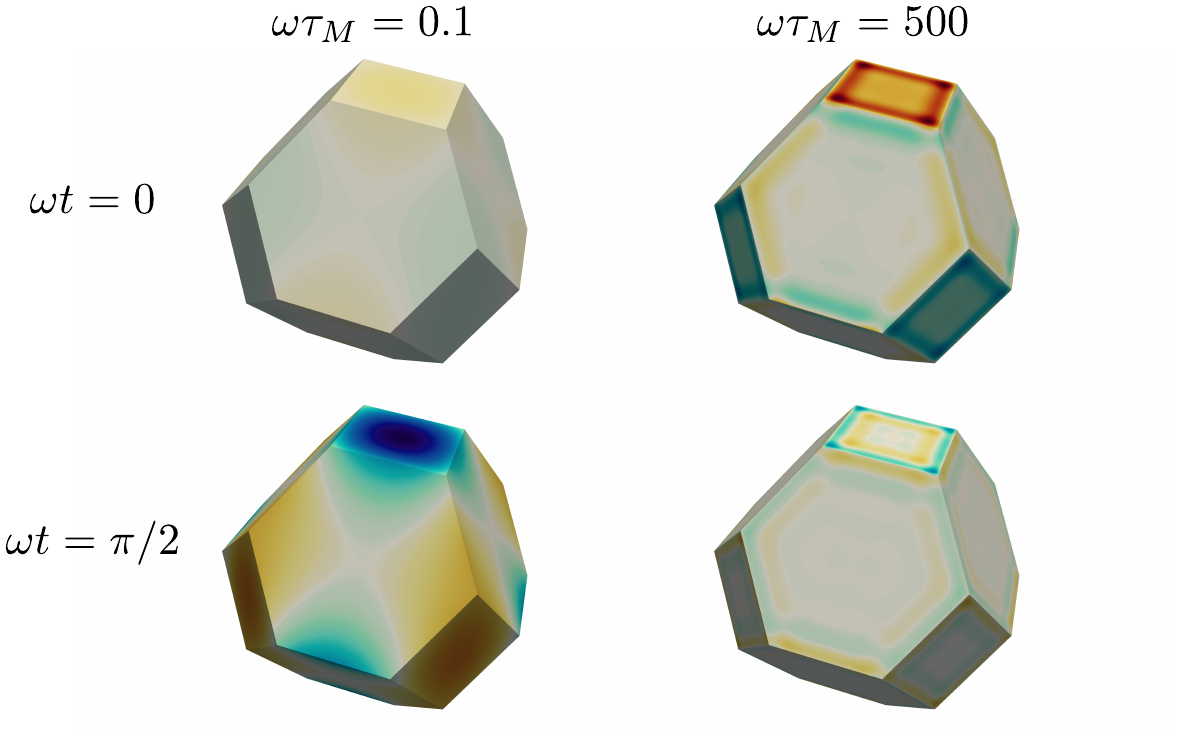}
 \caption{Images of boundary normal stress $\sigma_{nn}$ for 
time-harmonic deformation of frequency $\omega$ for tetrakaidecahedral grains. 
The 
imposed macroscale strain is proportional to $\cos \omega t$ where $t$ is 
time, and the principal axes of the deformation are aligned with the square 
faces which are at right angles to each other.  Orange 
colours are positive, blue colours are negative, and off-white is zero. Left 
column shows images at low dimensionless frequency $\omega \tau_M=0.1$; right 
column shows images at high dimensionless frequency $\omega \tau_M=500$. The 
top 
row is at time $t=0$ in the cycle, the bottom row at a quarter of a period 
later. The bottom left image is similar to that shown for steady-state creep in 
Figure 9 (networked) in \citep{Rudge2018}. At high dimensionless frequency
diffusion only influences a narrow boundary layer in the 
neighbourhood of the 
triple line (the lengthscale $l$ in \eqref{eq:ldef}); at low 
dimensionless frequency diffusion influences the whole grain 
boundary.}
 \label{fig:grains}
\end{figure}

A simple grain shape suitable for 3D calculations is the tetrakaidecadedron 
(truncated octahedron). This grain shape can tessellate 
three-dimensional space, and has 14 faces, 6 of which are squares and 8 of 
which are hexagons (\autoref{fig:grains}). Unlike the hexagonal geometry in 2D, 
for which any fourth 
rank tensor (like the compliance tensor) must be isotropic, the fourth rank 
tensor associated with a tessellation of tetrakaidecahedrons is in general 
anisotropic, although there are strong constraints provided by the cubic 
symmetry. With cubic symmetry the resistance to shear is described by two 
independent parameters.

The steady-state Coble creep viscosity tensor of a tiling of 
tetrakaidecahedrons was calculated in \cite{Rudge2018a} and expressed in terms 
of two shear viscosities $\eta^{(1)}$ and $\eta^{(2)}$,
\begin{equation}
 \eta^{(1)} = 0.0012242 \frac{k T d^3}{\delta D_{gb} \Omega}, \quad \eta^{(2)} 
= 
0.0034715 \frac{k T d^3}{\delta D_{gb} \Omega}, 
\end{equation}
where $\eta^{(1)}$ is the shear viscosity associated with deformation whose 
principal axes align with the square faces of the grain, and $\eta^{(2)}$ is 
associated with shear deformation at an angle of 45° to those axes. $d$ is 
defined as the distance between opposite square faces. One way of defining an 
average viscosity is in terms of a Voigt-average,
\begin{equation}
\eta^V = \frac{2}{5} 
\eta^{(1)} + \frac{3}{5} \eta^{(2)} = 0.0025726 \frac{k T d^3}{\delta D_{gb} 
\Omega}.
\end{equation}

\begin{figure}
\centering
 \includegraphics[width=0.605\columnwidth]{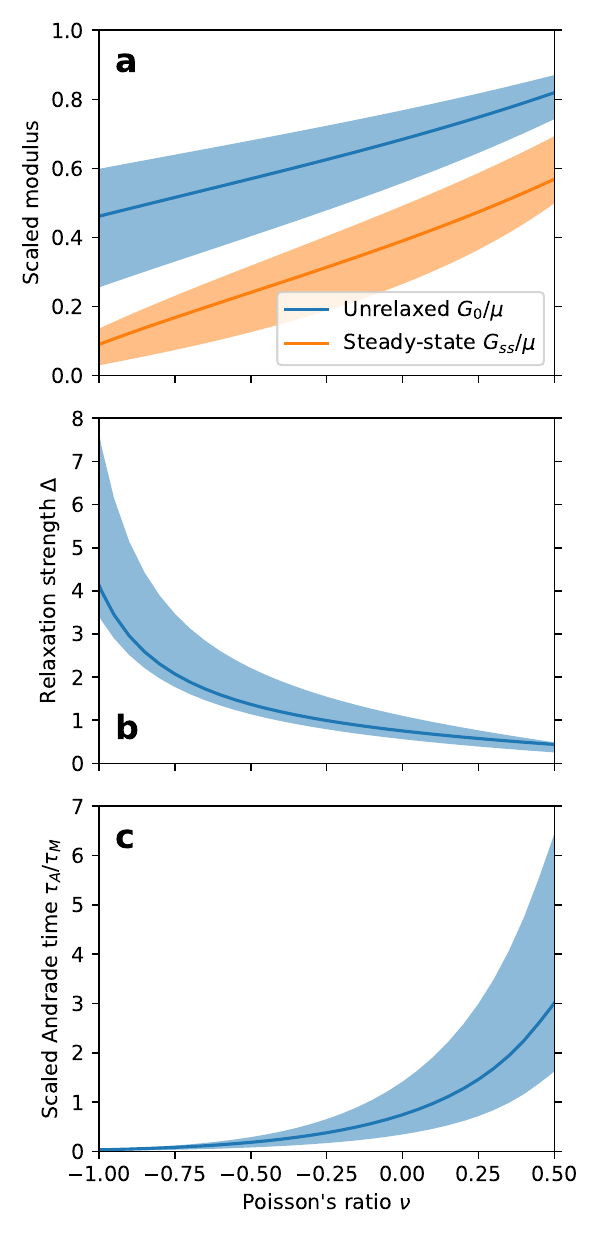}
 \caption{The same properties as \autoref{fig:hex_moduli}, but for a 
tessellation of tetrakaidecahedrons in 3D. The properties of such a tiling are 
anisotropic. Shaded bands are used to show how the parameters can vary with 
different orientations. The solid lines show results from Voigt-averaging 
the corresponding complex moduli.}
 \label{fig:tetra_moduli}
\end{figure}

\begin{figure}
 \includegraphics[width=\columnwidth]{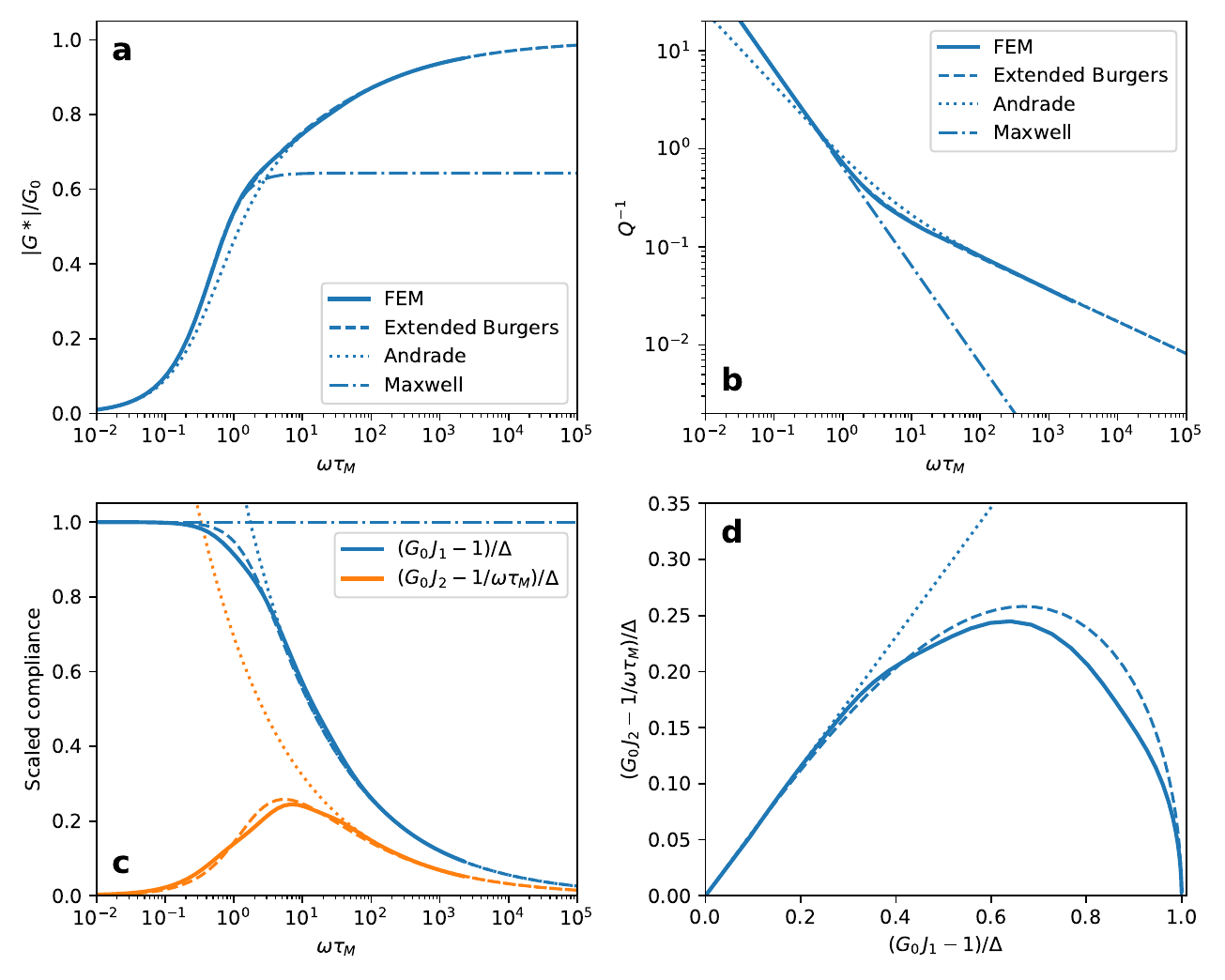}
 \caption{The same viscoelastic response functions as \autoref{fig:hex_all} but 
for the case of tetrakaidecahedral grains. The Poisson's ratio $\nu=0.3$ and 
the curves are shown for Voigt-averaged values.}
 \label{fig:tetra_all}
\end{figure}

The general viscoelastic response in shear can similarly be expressed in terms 
of two complex shear moduli as a function of frequency, or in terms of a 
particular average. \autoref{fig:tetra_moduli} presents the same model 
properties as given in \autoref{fig:hex_moduli}, but now for the case of 
tetrakaidecahedral grains. Due to the anisotropy, the results are presented in 
terms of bands of values, where the extremes are calculated in terms of the 
complex moduli associated with the directions for $\eta^{(1)}$ and 
$\eta^{(2)}$ above. The solid lines show the corresponding Voigt-average 
values (\autoref{sec:voigt}). \autoref{fig:tetra_all} shows a similar plot to 
\autoref{fig:hex_all} for the case of tetrakaidecahedrons.

The overall viscoelastic response for the tetrakaidecahedral grains is very 
similar to that for hexagonal grains, with small differences that can be 
expressed in 
terms of 
the different effective values of $\Delta$, $\tau_A/\tau_M$ and $\alpha$. The 
differences in $\alpha$ can be attributed to the differences in the angles at 
which grain boundaries meet at the triple lines. For a hexagon, the three 
angles are all 120°; for a tetrakaidecadedron the angles are 
125.26°-125.26°-109.47° ($125.26\text{°}= 
\operatorname{arccos}(-1/\sqrt{3})$, $109.47\text{°} = 
\operatorname{arccos}(-1/3)$). The Picu and Gupta \citep{Picu1996} stress 
exponent for these 
angles is $\lambda=\tfrac{1}{2}$ and hence $\alpha = \tfrac{1}{3}$, a little 
less than the value for hexagons. For a typical rock value of $\nu=0.3$, the 
relaxation 
strength $\Delta = 0.55$ and the scaled Andrade time is $\tau_A/\tau_M = 
1.7$. The most notable difference between \autoref{fig:hex_all} and 
\autoref{fig:tetra_all} is due to the change in relaxation 
strength, which is larger by a factor of 1.7 in \autoref{fig:tetra_all}. The 
high frequency behaviour is very similar as both the Andrade exponent and the 
ratio of 
Andrade time to Maxwell time are similar for the two geometries. 

\section{Discussion}\label{sec:disc}

\begin{table}[ht!]
\centering
 \begin{tabular}{cccl}
 \hline
  $\alpha$ & $\tau_A/\tau_M$ & $\Delta$ & Reference \\ \hline 
 0.3672092 & 1.31 & 0.330 & This study, hexagon model, $\nu=0.3$\\
 0.3333333 & 1.68 & 0.552 & This study, tetrakaidecahedron model, $\nu=0.3$ \\ 
\hline
0.3 & 11.1 & 0.202 & Lee et al. \citep{Lee2011} bicrystal model, sawtooth (S), 
$\varphi= 
30\text{°}$\\
0.3672092 & 248 & 0.0272 & Lee et al. \citep{Lee2011}, bicrystal model, 
truncated 
sawtooth (TS), $\varphi= 
60\text{°}$ \\
\hline
0.38 & 0.415 & - & Takei et al. \citep{Takei2014}, borneol\\
0.38 & 0.144 & - & Yamauchi and Takei \citep{Yamauchi2016}, borneol + 
diphenylamine 40, 41, 43\\
0.33 & 0.58 & 1.4 & Jackson and Faul \citep{Jackson2010}, synthetic olivine 
6585\\
0.274 & 0.124 & 1.04 & Jackson and Faul \citep{Jackson2010}, synthetic olivine 
6381, 
6585, 6365, 6261, 6328 \\
0.26 & 0.218 & 1.2 & Jackson \citep{Jackson2019}, synthetic olivine \\
0.217 & 1.63 & 0.76 & Qu et al. \citep{Qu2024}, synthetic dunite (SS-jacketed) 
A1802 \\
0.206 & 0.0299 & 1.76 & Barnhoorn et al. \citep{Barnhoorn2016}, synthetic MgO 
1096 
and 1077\\ 
0.247 & 0.216 & - & Priestley et al. \citep{Priestley2024}, fit to multiple 
data sets, 
$T_h=0.89$\\
0.228 & 0.0269 & - & Priestley et al. \citep{Priestley2024}, fit to multiple 
data sets, 
$T_h=1.0$\\\hline
 \end{tabular}
\caption{A comparison of the viscoelastic parameters determined in this study 
with some of those in the literature. $\alpha$ is the Andrade exponent, 
$\tau_A/\tau_M$ 
is the ratio of the Andrade time to the Maxwell time, and $\Delta$ is the 
relaxation strength. The synthetic olivine/ synthetic dunite samples also 
contain a small fraction of pyroxene grains. The fits of 
Priestley et al. \citep{Priestley2024} 
are functions of homologous temperature $T_h$, the ratio of the absolute 
temperature to the solidus temperature.}
 \label{tab:alphas}
\end{table}

\begin{figure}
\centering
\includegraphics[width=0.6\columnwidth]{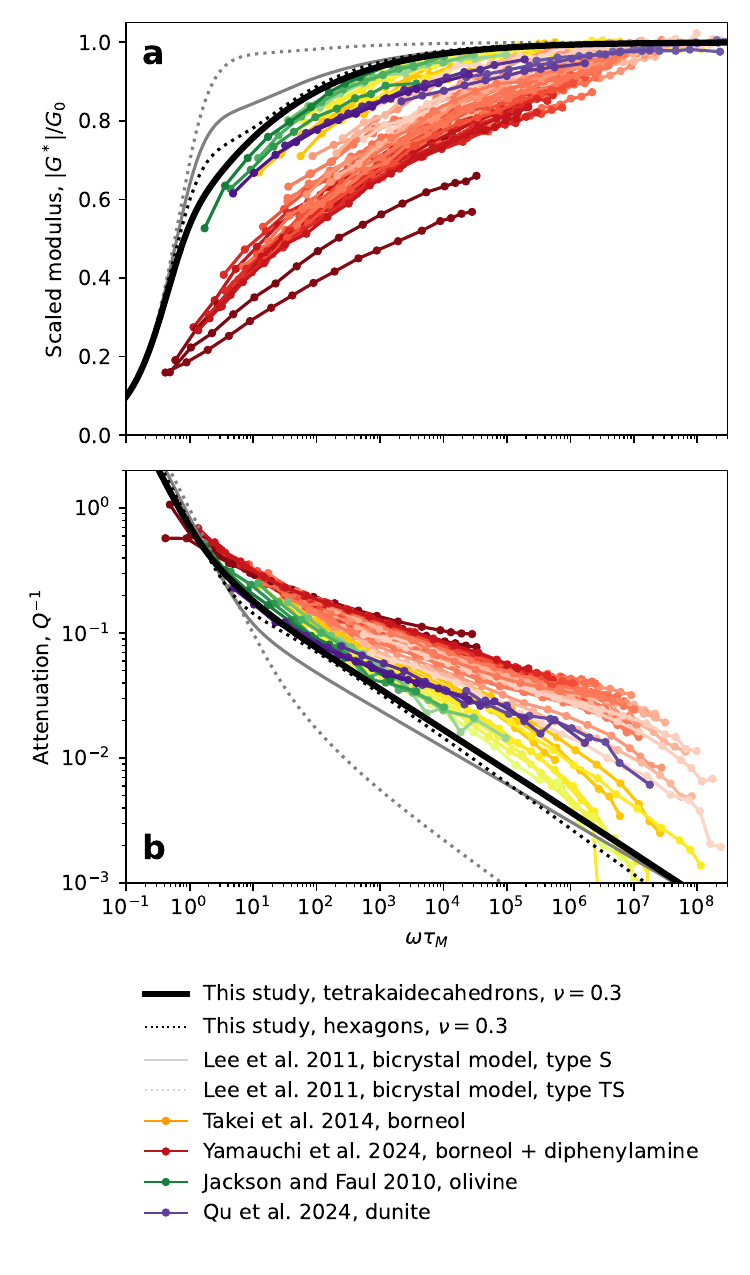}
 \caption{Comparison of the model presented here with literature estimates of 
a) scaled modulus and b) attenuation. Black lines show the model for 
tetrakaidecahedrons (solid line) and hexagons 
(dotted line) for a Poisson's ratio $\nu=0.3$. Results 
for the Lee et al. \citep{Lee2011} bicrystal model are shown as grey lines: the 
sawtooth 
(S) interface as a solid line (from their Figure 6), the truncated sawtooth (TS) 
interface as a dashed line (from their Figure 7). Yellow dots show borneol 
samples 27, 32, and 36 from Takei \citep{Takei2014}; red dots show borneol + 
diphenylamine samples 41, 42, and 50 from Yamauchi and 
Takei \citep{Yamauchi2024}. Shading is used 
to indicate temperature, with darker colours indicating high temperature (see 
Figure C1 and text of \citep{Yamauchi2024} for further details). Green dots 
show 
data for synthetic olivine polycrystal sample 6585 of 
Jackson and Faul \citep{Jackson2010} for 
temperatures between 1000~°C and 1200~°C. Purple dots show data for synthetic 
dunite polycrystal sample A1802 of Qu et al. \citep{Qu2024} between 1000~°C and 
1200~°C. 
Estimates of Maxwell time and unrelaxed modulus used for scaling laboratory data 
use those values given in the studies concerned.}
\label{fig:comparison}
\end{figure}

\autoref{tab:alphas} shows a comparison between the viscoelastic parameters 
calculated for the models here with those given in the literature, which have  
principally been obtained by fits to laboratory experiments. Also given in 
\autoref{tab:alphas} are parameters inferred from the 
bicrystal model of Lee et al. \citep{Lee2011}, determined by fitting the 
curves shown in 
their figures 6 and 7. \autoref{fig:comparison} provides a comparison between 
the models and laboratory data in plots of attenuation and scaled modulus.

The bicrystal model of \citep{Lee2011} describes the same physics as considered 
here, and differences only arise due to differences in geometry. Lee et al.
\citep{Lee2011} consider two types of interface between the two grains of the 
bicrystal: i) a sawtooth interface (S) and ii) a truncated sawtooth (TS) 
interface. 
These interfaces mimic those seen in an array of regular hexagons (referred to 
as the mode 2 and mode 1 directions in \citep{Raj1971}). One important 
difference between the bicrystal model and the present study is that in the 
bicrystal model each corner involves two sliding surfaces coming together, 
whereas at the triple junctions in the present study three sliding surfaces 
come together. The stress exponents are thus different in general (see 
\citep{Picu1996} Figure 5). For a 120° interior angle 
with two sliding surfaces
there are two values for $\alpha$; a value of 0.3672092 for eigenfields which 
are symmetric (which is relevant to the TS model, and identical to the array 
of hexagons here) and a value of 0.3 for eigenfields which are antisymmetric 
(relevant to the S model).

A more significant difference between the present study and that of Lee et al.
\citep{Lee2011} is the values of $\tau_A/\tau_M$ which are 
significantly larger for the bicrystal models of \citep{Lee2011} than the 
models here; that is to say the bicrystal models of \citep{Lee2011} show 
significantly less attenuation for the same dimensionless frequency (the grey 
lines are lower than the black lines on \autoref{fig:comparison}b). This 
difference must arise from the overall differences in 
grain geometry, 
where it should be noted that the bicrystal model has two lengthscales which 
define it rather than just one for the model here: there is one lengthscale 
associated with the variations of the interface, and another associated with 
the width of the bicrystal. Lee et al. \citep{Lee2011} present results where 
the width is 
5 times that of the scale of variations on the interface; it is expected that 
the ratio of these two lengthscales plays an important role in determining the 
amount of attenuation, as dissipation only takes place on the boundary but 
elastic energy is stored throughout the grains.

A number of laboratory studies have aimed to characterise the viscoelastic 
behaviour of rocks by fitting Andrade and extended Burgers models to forced 
oscillation data. The results of the present study are remarkably similar to 
those of Takei \citep{Takei2014} for polycrystalline borneol, which is an 
organic 
compound used as 
a rock analogue, for which laboratory experiments can be performed close to 
room 
temperature. Both the exponent $\alpha$ and the ratio $\tau_A/\tau_M$ are 
fairly close between the borneol studies and the model.  
Takei \citep{Takei2014} (and 
the 
subsequent Yamauchi and Takei \citep{Yamauchi2016,Yamauchi2024} studies) use an 
Andrade model for 
fitting the data, and so the total relaxation strength has not been estimated. 
As is the case in many experimental studies, data is not 
available at low enough frequencies (i.e. $\omega \tau_M \ll 1$) to 
accurately determine the 
relaxed modulus. In \autoref{fig:comparison}b which shows attenuation,
the yellow dots corresponding to the Takei \citep{Takei2014} study lie just 
above the 
model curve for tetrakaidecahedrons, demonstrating their similarity, 
particularly at low temperatures.

There are more significant differences between the present model and those 
based on laboratory experiments using synthetic 
rock aggregates deformed at high temperature 
\citep{Jackson2010,Jackson2019,Barnhoorn2016,Qu2024}. These studies provide 
evidence for somewhat lower values of the Andrade exponent $\alpha$ (as low as 
0.2). 
The studies also provide estimates of total relaxation strength $\Delta$ as 
they 
are fit 
with an extended Burgers model, although such estimates are not 
well-constrained due to the lack of low frequency data. The high frequency data 
are expected to constrain the Andrade parameters ($\alpha$ and $\tau_A$) and 
not $\Delta$. In the 
studies of 
\citep{Jackson2010,Barnhoorn2016,Jackson2019,Qu2024} the parameters of the 
extended Burgers 
model are not independently chosen: In \citep{Jackson2010} the high period 
cutoff is fixed at a specific value of $10^6$~s, and in 
\citep{Barnhoorn2016,Jackson2019,Qu2024} an assumption is made that the 
high 
period cutoff is at the Maxwell time, i.e. $\tau_H = \tau_M$. This assumption 
constrains 
the parameters in \autoref{tab:alphas} for these studies to satisfy 
$\tau_A/\tau_M = \left(\Gamma(1- \alpha) \Delta \right)^{-1/\alpha}$.  Note 
that in the extended Burgers models fits
for the hexagons $\tau_H= 0.166 \tau_M$ and for the tetrakaidecahedron 
$\tau_H= 0.685 \tau_M$, so the assumption that $\tau_H=\tau_M$ may be 
questioned.

\begin{figure}
\centering
 \includegraphics[width=0.7\columnwidth]{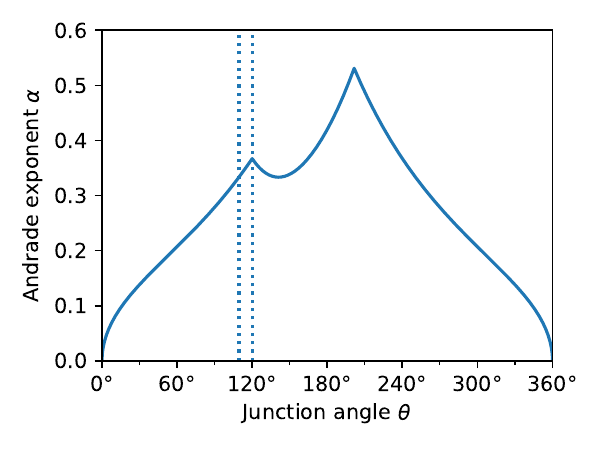}
 \caption{Andrade exponent $\alpha$ as a function of junction angle $\theta$ 
for a triple junction with angles $\theta$, $180\text{°}-\theta/2$ 
and $180\text{°}-\theta/2$. Dotted lines indicate the junction 
angles for the tetrakaidecahedron ($\theta=109.47\text{°}$, 
$\alpha=\tfrac{1}{3}$) and the hexagon ($\theta=120\text{°}$, 
$\alpha=0.3672092$).}
 \label{fig:picu_exponent}
\end{figure}

It is not clear why some laboratory studies show lower Andrade exponents than 
the models here. In the models, the Andrade exponent is purely controlled by 
the triple junction geometries, and significant departures from the symmetric 
120°-120°-120° junction angles are needed to produce Andrade exponents of 0.2. 
This is illustrated in \autoref{fig:picu_exponent}, which plots the Andrade 
exponent expected for asymmetric junctions with one angle of $\theta$ and two 
identical angles of $180\text{°}-\theta/2$. An exponent of 0.2 could be 
produced by having a $60\text{°}-150\text{°}-150\text{°}$ junction, but this 
seems far from what is observed in the microstructures. There may be some other 
source of anisotropy or geometrical feature that influences the exponent in 
some experiments, or there may be another dissipation mechanism producing 
attenuation that is mapped by the fitting to produce a lower effective Andrade 
exponent.

One additional mechanism of attenuation is that associated with 
elastically-accommodated grain boundary sliding. No explicit calculations of 
this mechanism have been made in the present work, but this has been considered 
in previous studies, notably for hexagons by Ghahremani \citep{Ghahremani1980} 
and in the 
bicrystal model by Lee et al. \citep{Lee2010,Lee2011}. Instead of 
assuming the grain boundaries are inviscid, one assumes there is a certain 
grain boundary viscosity $\eta_{gb}$ that describes the resistance to sliding. 
This grain boundary viscosity leads 
to an attenuation peak at a particular period $\tau_{e}$, which up to a 
dimensionless
prefactor is at
\begin{equation}
 \tau_{e} = \frac{\eta_{gb} d}{\mu \delta}.
\end{equation}
The relaxation strength for this mechanism can be read off from 
\autoref{fig:hex_moduli}a and \autoref{fig:tetra_moduli}a as
\begin{equation}
 \Delta_{e} = \frac{\mu}{G_0} - 1
\end{equation}
For $\nu=0.3$ the hexagon model yields $\Delta_{e}=0.233$ and the 
tetrakaidecadedron model yields $\Delta_{e}=0.310$. 
Ghahremani \citep{Ghahremani1980} 
have shown that the viscoelastic response of this mechanism is well-described 
by a Burgers model with relaxation at a single period. Somewhat smaller 
relaxation strengths were found for the bicrystal model of Lee et al.
\citep{Lee2011,Lee2010} with $\Delta_{e}=0.098$ for the S model and 
$\Delta_{e}=0.080$ for the TS model.

The relevance and importance of elastically-accommodated grain-boundary sliding 
in experimental studies is debated. Many studies fit observations by including 
a high frequency attenuation peak, which some ascribe to 
elastically-accommodated grain boundary sliding (e.g. \citep{Qu2024}) and 
others 
do not (e.g. \citep{Takei2014, Yamauchi2024}). As noted by 
Takei \citep{Takei2014} 
(their equation (22)) and Jackson \citep{Jackson2002} (their equation (13)), a 
simple model of grain boundary viscosity by Ashby
\citep{Ashby1972} would suggest that the ratio $\tau_e/\tau_M$ should be around 
$b^2/d^2$ where $b \sim 10^{-10}$~m is the atomic size and $d$ is the grain 
size. For 1~$\mu$m grains, $\tau_e/\tau_M \sim 10^{-8}$; for 10~$\mu$m grains, 
$\tau_e/\tau_M \sim 10^{-10}$; and for 1~mm grains $\tau_e/\tau_M \sim 
10^{-14}$. Thus any peak from elastically-accommodated 
grain boundary sliding needs data at very high frequencies relative to the 
Maxwell time to be resolved. The peak that Qu et al. \citep{Qu2024} report has 
$\tau_e/\tau_M = 7 \times 10^{-6}$ and a relaxation strength $\Delta_e = 0.012$. 
This is a significantly lower relaxation strength than the simple models here 
predict. The differences in relaxation strength might be due to 
geometrical factors not modelled here, such as heterogeneity in grain size in 
real materials \citep{Lee2010, Jackson2014}. 

The studies of \citep{Takei2014,Yamauchi2016,Yamauchi2024} on borneol have 
suggested there is a significant high frequency dissipation peak around  
$\tau_P/\tau_M = 6 \times 10^{-5}$ that cannot be attributed to 
elastically-accommodated grain boundary sliding because of the independence of 
the peak position on grain size \citep{Takei2014}. The peak has been modelled as 
scaling on the Maxwell time in the same way as for 
diffusionally-assisted/accommodated grain boundary sliding, but with a peak 
width and height that varies with homologous temperature. In particular the 
relaxation strength $\Delta_p$ varies from $\Delta_p=0.1$ at a homologous 
temperature of $T_h=0.92$ to $\Delta_p=0.5$ at a homologous temperature of 
$T_h=1.0$. The evidence for this peak is seen in the attenuation data in 
\autoref{fig:comparison} as an increase in attenuation with increasing 
homologous temperature (the darker red dots). There is as yet no 
micromechanical model which explains this peak, which has tentatively been 
suggested to arise from changes 
associated with premelting of grain boundaries as the material approaches the 
melting point \citep{Yamauchi2016,Yamauchi2024}. A notable challenge for future 
micromechanical models is to explain and quantify this peak.

Understanding the 
viscoelastic behaviour of rocks at high homologous temperatures is of particular 
relevance to Earth sciences when interpreting the significant variations in 
seismic wavespeed, as well as increased attenuation, that occur near the 
lithosphere-asthenosphere boundary 
\citep[e.g.][]{Ma2020,Priestley2024}. The models here do not produce enough 
attenuation to explain what is seen in seismology. For example, with a 
steady-state asthenospheric shear viscosity $\eta = 2 \times 10^{19}~$Pa~s 
and an intrinsic shear modulus $\mu = 70$~GPa, the Maxwell time $\tau_M = 3 
\times 10^{8}$~s, i.e. around 10 years. For a surface wave with a period around 
$T=50~$s, the dimensionless frequency is $\omega \tau_M = 2 \pi \tau_M/T = 4 
\times 10^{7}$, for which the models yield $Q^{-1} \sim 10^{-3}$ 
(\autoref{fig:comparison}), significantly less than the asthenospheric $Q^{-1} 
\sim  10^{-2}$ estimated by Ma et al. \citep{Ma2020} and given in PREM 
\citep{Dziewonski1981}. Previous work in extrapolating laboratory experiments 
by Jackson and Faul \citep{Jackson2010} suggested that 
diffusionally-accommodated grain 
boundary sliding can account for the $Q^{-1} 
\sim 10^{-2}$ level of attenuation at seismic frequencies. This 
extrapolation was based on a proposed variation of Maxwell time as linear 
with grain size, $\tau_M \propto d$, rather than $\tau_M \propto d^3$ as 
expected from theory for grain-boundary diffusion. More recent experiments in 
the same laboratory \citep{Qu2022} have since observed a much stronger grain 
size dependence with $\tau_M \propto d^{3.17}$, closer to the behaviour 
predicted by theory, and for which extrapolation does not yield sufficient 
attenuation at seismic frequencies. Thus additional mechanisms of dissipation, 
such as those associated with premelting \citep{Yamauchi2016,Yamauchi2024}, 
impurities \citep{Takei2022},
elastically-accommodated grain boundary sliding \citep{Karato2012}, or 
dislocations \citep{Sasaki2019,Hein2025} are needed to explain the seismic 
observations.

There are a number of natural ways the present modelling work can be extended. 
One 
natural extension is to consider diffusion within the interior of grains rather 
than along the grain boundaries. An example of such a calculation for hexagons 
is given in \autoref{sec:vdiff}, and the most notable result is that the 
relationship between the Andrade exponent and the stress singularity exponent 
changes to $\alpha = 1 - \lambda$, such that $\alpha=1/2$ is expected for 
tetrakaidecahedral grains. A key flexibility of the finite element modelling 
approach taken here is that can be adapted to model a wide range of 
different grain shapes with 
potentially anisotropic properties. Calculations could be performed using a 
synthetic 
rock texture which has a 
range of grain sizes, e.g. using Neper \citep{Quey2018}. The expectation would 
be that the Andrade exponent would remain determined by the triple junction 
angles, but that the effective $\tau_A/\tau_M$ and $\Delta$ parameters may 
change. All the calculations here are for materials in solid-state; solid 
both in the grain interiors and on the grain boundaries. The effect 
of melt and melt geometry \citep{Rudge2018} could be 
explored, as was done for the case of steady-state diffusion creep viscosity in 
\citep{Rudge2018a}. The presence of melt allows for a 
steady-state bulk 
viscosity of the material, and so calculations with melt would allow for both 
shear and bulk attenuation phenomenon to be explored. However, perhaps most 
interesting would be to consider the temperature dependent properties of the 
grain boundaries and the triple lines and explore the effect of premelting and 
sub-solidus mechanisms of attenuation more generally.

\section{Conclusions}\label{sec:conc}

We have presented a simple grain-scale model of 
diffusionally-accommodated/-assisted grain-boundary sliding and calculated its 
macroscale 
viscoelastic rheology. The rheology can be well-described by an extended 
Burgers model with a power-law distribution of relaxation times and a long 
period cut-off. The parameters of this extended Burgers model are in remarkable 
agreement with those inferred from laboratory experiments on the 
rock-analogue material borneol. The Andrade exponent $\alpha$ is around a 
third, the Andrade time $\tau_A$ is comparable to the Maxwell time $\tau_M$, 
and the relaxation strength $\Delta$ is around 0.6. The 
close agreement between the laboratory experiments and the model gives 
confidence that the model can be 
used to describe the viscoelastic behaviour of a wide range of 
polycrystalline materials in the regimes of 
diffusionally-accommodated/-assisted grain-boundary sliding. However, laboratory 
experiments also indicate that there remain additional mechanisms of attenuation 
that are still  to 
be fully understood, particularly as the temperature of the material approaches 
the 
melting point.

\clearpage

\appendix
\renewcommand\theequation{\thesection\,\arabic{equation}}

\section{Governing equations}\label{sec:govern_eq}

\subsection{Elasticity}

Grain interiors are assumed to behave linear elastically, with conservation 
of momentum
\begin{equation}
 \nabla \cdot \tensor{\sigma} = \mathbf{0} \label{eq:elasticity}
\end{equation}
for a symmetric stress tensor $\tensor{\sigma}$, related to the displacement 
$\u$ by the constitutive law
\begin{equation}
 \sigma_{ij} = \lambda e_{kk} \delta_{ij}  + 2 \mu e_{ij} 
\end{equation}
where $\lambda$ and $\mu$ are the Lame moduli and $\tensor{\e}$ is the 
symmetric strain tensor,
\begin{equation}
 e_{ij} = \frac{1}{2} \left( \pdd{u_i}{x_j} + \pdd{u_j}{x_i} \right).
\end{equation}

\subsection{Grain boundary diffusion}

On the boundaries between grains the elastic displacement is discontinuous. 
Diffusion of atoms and vacancies along the grain boundaries leads to plating 
out or removal of material. Grain boundary diffusion is described by 
\citep[e.g.][section 2.2]{Rudge2018a}
\begin{equation}
 \left[ \dot{\u} \cdot \n \right] + \Omega \delta D^\text{gb}_\text{v} 
\nablas^2 c = 0 \label{eq:bdry_diff}
\end{equation}
where square brackets are used to denote the jump in a quantity across the 
grain boundary. $\dot{\u} \equiv \partial \u / \partial t$ is the velocity, the 
rate of change of the displacement. $\n$ is a unit vector normal to the 
grain boundary. $\left[ \dot{\u} \cdot \n \right]$ 
represents the jump in normal velocity across the grain boundary, which is the 
rate at which new material is plated out on the grain boundary. This plating is 
accommodated by diffusion, where $D^\text{gb}_\text{v}$ is the vacancy 
diffusivity along the grain boundary, $\delta$ is the grain boundary width, 
$\Omega$ is the atomic volume, and $c$ is the concentration of vacancies. 
$\nablas$ represents the gradient operator restricted to the grain boundary; 
$\nablas^2$ is the surface Laplacian operator.  

The concentration of vacancies is related to the normal stresses on the 
boundary by Herring's relation \citep{Herring1950}
\begin{equation}
 c = c_0 \left( 1 + \frac{\Omega p }{k T}   \right) \label{eq:convac}
\end{equation}
where $c_0$ is the equilibrium concentration of vacancies, $k$ is Boltzmann's 
constant and $p$ is the normal component of the traction on the boundary,
\begin{equation}
p \equiv \n \cdot \stress \cdot \n \quad \text{on }S, \label{eq:pdef}
\end{equation}
where $S$ is the grain-boundary interface. \eqref{eq:convac} can be substituted 
into \eqref{eq:bdry_diff} to write
\begin{equation}
 \left[ \dot{\u} \cdot \n \right] + \frac{\Omega \delta D^\text{gb}}{k T}
\nablas^2 p = 0 \label{eq:bdry_diff3}, \\
\end{equation}
where $D^\text{gb} = D^\text{gb}_\text{v} c_0$ is the self-diffusion 
coefficient 
for grain boundary diffusion.

\subsection{Grain boundary sliding}

The simplest model of grain boundary sliding is to consider the interface 
between two grains as if it were coated with a viscous fluid that offers 
resistance 
to being sheared,
\begin{equation}
 \n \times \tensor{\sigma} \cdot \n = \frac{\eta_\text{gb}}{\delta} [\n \times 
\dot{\u}] \quad \text{ on }S \label{eq:gbs}
\end{equation}
where $\eta_\text{gb}$ is the grain boundary viscosity and $\delta$ is the 
grain boundary width. In the present work, we focus on timescales long enough 
such that the shear stresses on the grain boundaries are 
relaxed, and thus impose
\begin{equation}
 \n \times \tensor{\sigma} \cdot \n = \mathbf{0} \quad \text{on }S. 
 \label{eq:gbs_simple}
\end{equation}

Together \eqref{eq:elasticity}, \eqref{eq:pdef}, \eqref{eq:bdry_diff3} and 
\eqref{eq:gbs_simple}  describe a coupled problem to be solved for the 
displacement $\u$ inside the grains and the normal stresses $p$ on the 
interfaces between grains. 

\section{Laplace transform}

A standard approach to studying the viscoelastic response of a material is to 
Laplace transform 
the variables in time as
\begin{equation}
 \tilde{f} (s) = \int_0^\infty f(t) \e^{- s t} \; \d t
\end{equation}
where tildes are used to denote Laplace-transformed variables. Under this 
transform differentiation in time becomes multiplication by $s$, assuming 
initial conditions are such that all variables vanish before time 0 i.e. before 
the 
application of some forcing (an assumption of causality). Laplace transforming 
replaces $\dot{\u}$ by $s 
\tilde{\u}$. In the sections that follow we work with Laplace transformed 
variables, but for clarity of notation we do not explicitly write tildes above 
each Laplace-transformed variable.

\section{Weak forms}

For finite element analysis the governing equations are most usefully written 
in weak form. Consider a domain $V$ potentially containing many 
individual grains. Taking the dot product of the complex conjugate of 
\eqref{eq:elasticity} with a vector test function $\v$, integrating over the 
domain $V$ and using Green's identity yields
\begin{equation}
 \int_V \tensor{\sigma}^{\u*} : \tensor{e}^{\v} \;\d V - \int_{S_i} p^* [\v 
\cdot 
\n]  
\;\d S  = \int_{S_0} \t^{0*} \cdot \v \; \d S  \label{eq:weak_elastic}
\end{equation}
where $S_i$ represents grain boundary surfaces; $S_0$ represents any 
exterior surfaces; and $\t^{0}$ any imposed tractions on those exterior 
surfaces. $\tensor{\sigma}^{\u}$ is the stress tensor formed from the 
displacement $\u$. $\tensor{e}^{\v}$ is the symmetric strain tensor formed from 
the test function $\v$. The colon ($:$) denotes the tensor contraction on two 
indices. The terms involving the normal and tangential components of the 
traction at grain boundaries have been rewritten using \eqref{eq:pdef} and 
\eqref{eq:gbs_simple}. Superscript $^*$ denotes complex conjugation. Note that 
both the 
displacement and the test function are complex functions, and the 
Laplace transform parameter $s$ 
may also be complex.

Multiplying the complex conjugate of \eqref{eq:bdry_diff3} with a scalar test 
function $q$, integrating over the grain boundary surfaces and using Green's 
identity yields
\begin{equation}
-\int_{S_i} [\u^* \cdot 
\n]  q
\;\d S + \frac{\Omega \delta D^\text{gb}}{k T s^*} \int_{S_i} \nablas p^* \cdot 
\nablas q 
\; \d S  =  -\frac{\Omega \delta }{s^*}  \int_C 
{\mathbf{j}_\text{v}^{*0}} \cdot \boldsymbol{\nu} \,q \; \d l 
\label{eq:weak_diff}
\end{equation}
where $C$ represents the triple lines where grain boundary surfaces intersect
and $\boldsymbol{\nu}$ is an outward-pointing conormal at the triple line 
(i.e. a vector perpendicular to both the grain-boundary normal and a vector in 
the direction of the contact line). $\mathbf{j}^0_\text{v}$ is the flux of 
vacancies into the triple line. It will be assumed that the
integral over $C$ vanishes, which is an assumption that there is no net flux 
into or out of the 
triple lines. 

\section{Periodic homogenisation}

Consider a collection of identical grains which tessellate space. The 
displacement within each grain can be decomposed as
\begin{equation}
 \u = \U + \hat{\u} \label{eq:decom}
\end{equation}
where $\U$ represents a rigid body motion and $\hat{\u}$ represents 
the remaining elastic deformation. Each grain in the tessellation will be 
assumed to have the same elastic deformation $\hat{\u}$, but that $\U$ will 
vary in a systematic and prescribed manner. Following \citep{Rudge2021} 
(section 3), the 
kinematics of the rigid body motions will be prescribed in terms of macroscale 
strain measures $\tensor{\Gamma}$ and $\tensor{K}$, where the symmetric part of 
$\tensor{\Gamma}$ represents the macroscale strain tensor. The jump in the 
rigid body displacement across the grain boundaries is 
\citep[][equation (21)]{Rudge2021}
\begin{equation}
 \left[\U \right] = \tensor{\Gamma} \cdot \R + \left(\tensor{K} \cdot \R 
\right) \times \vector{d}
\end{equation}
where $\R$ is a vector 
which joins the centroids of the neighbouring grains, and $\vector{d}$ is a 
vector joining the centroid of the neighbouring grain to the position on the 
grain boundary \citep[][Figure 1]{Rudge2021}. In the present work we will 
only consider upscaling to a Cauchy 
continuum, not a more general micropolar continuum as was done in 
\citep{Rudge2021}. We will thus neglect the strain 
measure $\tensor{K}$ associated with gradients in microrotation, prescribing
\begin{equation}
\left[\U \cdot \n \right] = \n \cdot \tensor{\Gamma} \cdot \R,  \label{eq:U1}
\end{equation}
and solve for a vector displacement field $\hat{\u}$ which is identical in each 
grain in the periodic tessellation. Substitution of \eqref{eq:decom} and 
\eqref{eq:U1} into the weak forms \eqref{eq:weak_elastic} and 
\eqref{eq:weak_diff} yields
\begin{gather}
 \int_V \tensor{\sigma}^{\hat{\u}*} : \tensor{e}^{\v} \;\d V - \int_{S_i} p^* 
[\v 
\cdot 
\n]  
\;\d S  = 0 \label{eq:weak_elastic2}\\
- \int_{S_i} [\hat{\u}^* \cdot 
\n]  q
\;\d S + \frac{\Omega \delta D^\text{gb}}{k T s^*} \int_{S_i} \nablas p^* \cdot 
\nablas q 
\; \d S = \int_{S_i} \left(\n \cdot \tensor{\Gamma}^* \cdot \R \right) q
\;\d S 
\label{eq:weak_diff2}
\end{gather}
which given $\tensor{\Gamma}$ can be solved for $\hat{\u}$ and $p$.

\subsection{Equivalent energetics}

In upscaling, we wish to replace the discrete collection of grains by an 
equivalent continuum description, with a macroscale effective stress tensor 
$\overline{\tensor{\sigma}}$ related to the macroscale strain measure 
$\tensor{\Gamma}$ by
\begin{equation}
\overline{\sigma}_{ij} =  C_{ijkl} \Gamma_{kl} \label{eq:stress_strain}
\end{equation}
We determine the effective stiffness tensor $C_{ijkl}$ by demanding that both 
the grain-scale model and the upscaled continuum have the same energetics, i.e. 
they store and dissipate energy at the same rate \citep{Rudge2021}. This is 
achieved by 
demanding equality between the microscale
\begin{equation}
 \Psi = \frac{1}{V} \left(  \int_V \tensor{\sigma}^{\u*} : \tensor{e}^{\u} \;\d 
V 
 - \frac{\Omega \delta 
D^\text{gb}}{k T s^*} \int_{S_i} \nablas p^* \cdot 
\nablas p 
\; \d S  \right) \label{eq:diss}
\end{equation}
and the macroscale
\begin{equation}
 \Psi = \overline{\sigma}^*_{ij} \Gamma_{ij} = \Gamma_{ij} C_{ijkl}^* 
\Gamma_{kl}^*.
\end{equation}

\section{Scaling}

The equations can be simplified further by scaling. There is a natural 
lengthscale associated with the size of the grains, which we label $d$. If 
$u_0$ is a typical displacement, then a natural scale for stresses is $\mu u_0 
/ d$ where $\mu$ is the intrinsic elastic shear modulus of the grains. 
Importantly, there is a natural timescale, given by
\begin{equation}
 \tau = \frac{k T d^3}{\delta D^\text{gb} \Omega \mu}
\end{equation}
which up to a numerical factor is a Maxwell time for the material, a ratio of 
the steady-state Coble creep viscosity to the shear modulus. Once the governing 
equations have been scaled, there are only two dimensionless parameters which 
remain to describe the rheology: the dimensionless Poisson's ratio $\nu$ and 
the dimensionless Laplace transform parameter $s \tau$ (or equivalently the 
dimensionless frequency $\omega \tau$ of oscillation when considering time 
harmonic loading).

\section{Numerical Implementation}\label{sec:num_imp}

The computational domain consists of a single grain. A finite element mesh of 
the domain is constructed (triangular in 2D, tetrahedral in 3D). The variable 
$\hat{\u}$ is solved for throughout the domain, the variable $p$ solved for 
only on the exterior boundary of the domain. Periodic boundary conditions on 
$p$ 
are applied to impose invariance under translations by lattice 
vectors, where a point on each grain boundary is identical under translation to 
a point on the opposite face. Since the grain boundaries are on the edges of 
the computational domain, each section of domain boundary contains half the 
total grain boundary, but by periodicity the same boundary is traversed twice 
when integrating round the total surface of the grain. This simplifies the 
calculation of quantities involving jumps across the interface, as a single 
integral on a grain boundary involving a jump can be split into two separate 
integrals only involving displacements on one side of the boundary. The weak 
forms \eqref{eq:weak_elastic3} and \eqref{eq:weak_diff3} can then be written as
\begin{align}
 \int_{V_\text{grain}} \tensor{\sigma}^{\hat{\u}*} : \tensor{e}^{\v} \;\d V - 
\int_{S_\text{grain}} p^* 
\v 
\cdot 
\n
\;\d S  &= 0 \label{eq:weak_elastic3}\\
-\int_{S_\text{grain}} \hat{\u}^* \cdot 
\n \, q
\;\d S + \frac{\Omega \delta D^\text{gb}}{2 k T s^*} \int_{S_\text{grain}} 
\nablas p^* \cdot 
\nablas q 
\; \d S &= \frac{1}{2} \int_{S_\text{grain}} \left(\n \cdot \tensor{\Gamma}^* 
\cdot \R \right) q
\;\d S 
\label{eq:weak_diff3}
\end{align}
where the factors of $1/2$ come from accounting for traversing the same 
boundaries twice as one integrates round a whole grain. 

The weak form was discretised using a P2 vector element for $\hat{\u}$ and a P2 
scalar element for $p$. The resulting set of linear equations were solved 
iteratively using MINRES and a GAMG preconditioner. The weak forms were 
discretised using the dolfinx python library \citep{BarattaEtal2023} with 
linear algebra performed using PETSc \citep{petsc-web-page}.

\begin{figure}
\centering
 \includegraphics[width=0.7\columnwidth]{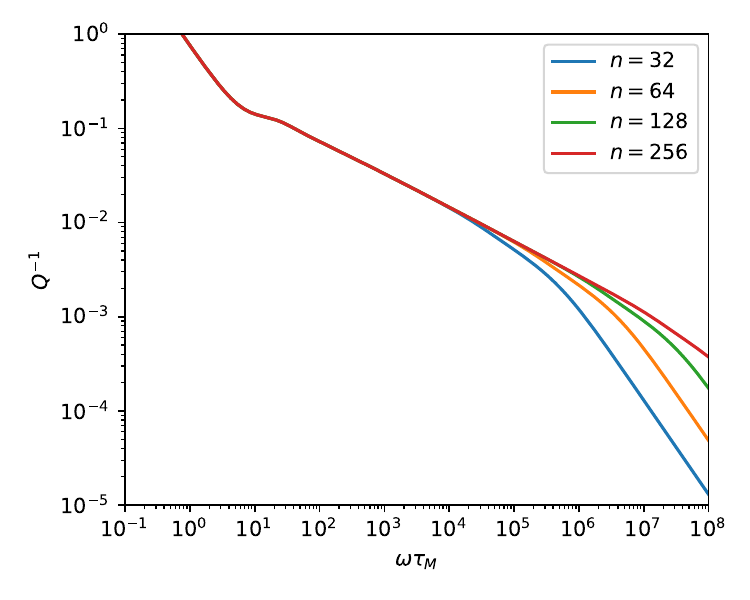}
 \caption{Influence of mesh resolution on finite element calculation of 
attenuation. The attenuation $Q^{-1}$ is plotted against the normalized 
frequency $\omega \tau_M$ for hexagonal grains for different values of $n$, the 
number of mesh cells on each edge of the hexagon.}
\label{fig:resolution}
\end{figure}

The accuracy and precision of the numerical results depends on mesh resolution. 
The effect of changing resolution is illustrated in \autoref{fig:resolution} 
which shows calculations of attenuation for hexagonal grains. The mesh 
of the regular hexagon is made up of cells of equilateral triangles of equal 
size, and the resolution specified by a parameter $n$ which gives the number 
of mesh cells per edge. Increasing $n$ by a factor of 2 increases the total 
number of mesh cells by a factor of 4. Simulations shown in the main text use 
$n=256$ which has 393,216 triangular mesh cells. As can be seen in 
\autoref{fig:resolution} mesh resolution limits the highest frequencies that 
can 
be resolved, where the attenuation is underestimated at high frequencies when 
the resolution is too coarse. This is as expected from equation 
\eqref{eq:ldef}, where diffusion acts over a length scale $l \propto 
\omega^{-1/3}$ at high frequencies, and this scale must be well-resolved to get 
accurate estimates of attenuation. Each doubling of $n$ leads to an 
eightfold-increase in frequency that can be resolved.

Producing well-resolved simulations in three-dimensions is computationally 
much more demanding than in two-dimensions. The simulations here were performed 
at a range of resolutions up to $n=32$ mesh cells per edge of the 
tetrakaidecadedron, with a total of $2,033,250$ tetrahedral mesh cells. Results 
shown in the main text (\autoref{fig:tetra_moduli}) are limited to frequencies 
lower that $\omega \tau_M<2 \times 10^3$ to be restricted to a range of 
frequencies that are well resolved.

\section{Viscoelastic response functions}

The main outcome of the finite element modelling is the tensor $C_{ijkl}$ which 
relates macroscale stress and strain in the Laplace-transformed variables of 
\eqref{eq:stress_strain}. For 
the simplest cases, this response is isotropic and can be written
\begin{equation}
 C_{ijkl} = s \tilde{K}(s) \delta_{ij} \delta_{kl} + s \tilde{G}(s) 
\left(\delta_{ik} \delta_{jl} + \delta_{il} \delta_{jk} - \frac{2}{N} 
\delta_{ij} \delta_{kl} \right)
\end{equation}
where $N$ is the dimension of the space, and $\tilde{K}(s)$ and $\tilde{G}(s)$ 
are the Laplace transforms of the bulk and shear relaxation moduli. The factors 
of $s$ in the definitions above arise from the standard definition of $G(t)$ 
in the time domain as the stress response to a unit step of strain. Knowledge 
of the functions $\tilde{K}(s)$ and $\tilde{G}(s)$ completely describes the 
response of the 
material, but depending on the type of loading considered other forms of
response function may be preferred. For example, the strain response to a unit 
step of stress is termed the creep compliance and denoted $J(t)$ in the time 
domain. The Laplace-transformed variables satisfy the reciprocal relation
\begin{equation}
 s \tilde{J}(s) = \frac{1}{s \tilde{G}(s)}.
\end{equation}
Detailed discussion of the conversions 
between different sets of viscoelastic response function can be found in 
textbooks (e.g. \citep{Nowick1972,Tschoegl1989}).

One of the most common experimental tests of viscoelastic behaviour is a forced 
oscillation test, where the sample has a time-harmonic deformation of a 
given frequency $\omega$ imposed. The stress response to a unit time harmonic 
strain is the complex shear modulus, given by
\begin{equation}
 G^*(\omega) = \left . s \tilde{G}(s) \right|_{s=i \omega}.
\end{equation}
The 
corresponding strain response to unit time harmonic stress is the complex shear 
compliance, given by
\begin{equation}
 J^*(\omega) = \left . s \tilde{J}(s) \right|_{s=i \omega}
\end{equation}
where there is the reciprocity $J^*(\omega) G^*(\omega) = 1$. The 
superscript * on $J^*(\omega)$ and $G^*(\omega)$ does not denote complex 
conjugation, but is a standard notation used for these quantities.  As complex 
quantities, $G^*(\omega)$ and $J^*(\omega)$ can be written in terms of real and 
imaginary parts as
\begin{align}
 G^*(\omega) &= G_1(\omega) + i G_2(\omega),\\
 J^*(\omega) &= J_1(\omega) - i J_2(\omega).
\end{align}
The reciprocal quality factor or loss tangent is
\begin{equation}
Q^{-1}(\omega) = \tan \arg G^*(\omega)  = \frac{G_2(\omega)}{G_1(\omega)} = 
\frac{J_2(\omega)}{J_1(\omega)}.
\end{equation}

\section{Response functions for simple models of 
viscoelasticity}\label{sec:simplemodels} 

For reference we provide here the response functions for some standard models 
of viscoelasticity. In each case $J_0$ refers to the unrelaxed compliance.

\subsection{Maxwell}\label{sec:maxwell}

The Maxwell model is represented by a spring and dashpot connected in series 
and has compliance
\begin{equation}
 s \tilde{J}(s) = J_0 \left(1 + \frac{1}{s \tau_M} \right)
\end{equation}
where $\tau_M$ is the Maxwell time. In the time domain the corresponding 
creep function is
\begin{equation}
 J(t) = J_0 \left( 1+ \frac{t}{\tau_M} \right).
\end{equation}

\subsection{Kelvin-Voigt}

The Kelvin-Voigt model is represented by a spring and dashpot connected in 
parallel and has compliance and creep function
\begin{align}
s \tilde{J}(s) &= \frac{J_r}{1 + s \tau_r},\\
J(t) &= J_r \left(1 - \e^{-t/\tau_r} \right),
\end{align}
where $\tau_r$ is the relaxation time and $J_r$ is the relaxed compliance. 
\subsection{Burgers}

The Burgers model combines a Maxwell model and a Kelvin-Voigt model in series, 
with compliance and creep function
\begin{align}
 s \tilde{J}(s) &= J_0 \left({1 + \frac{\Delta}{1 + s \tau_r} + \frac{1}{s 
\tau_M}} 
\label{eq:burgers} \right), \\
J(t) &= J_0 \left( 1 + \Delta \left(1 - \e^{-t/\tau_r} \right) + 
\frac{t}{\tau_M} \right),
\end{align}
where $\Delta$ is the anelastic relaxation strength.

\subsection{Andrade}\label{sec:andrade}

The Andrade model \citep{Andrade1910} has compliance and creep function
\begin{align}
 s \tilde{J}(s) &= J_0 \left({1 + \frac{\Gamma(1+ \alpha)}{(s \tau_A)^\alpha} 
+ \frac{1}{s \tau_M}}
\right), \\
J(t) &= J_0 \left(1 + \left(\frac{t}{\tau_A} \right)^\alpha + \frac{t}{\tau_M} 
\right),
\end{align}
where $\alpha$ is a dimensionless Andrade exponent, $\tau_A$ is the Andrade 
timescale 
and 
$\Gamma(z)$ is the Gamma function.

\subsection{Extended Burgers}\label{sec:eb}

The extended Burgers model has
\begin{align}
 s \tilde{J}(s) &= J_0 \left({1 + \Delta \int_0^\infty \frac{\gamma 
\rho(\gamma)}{s+ \gamma} \; \d \gamma + \frac{1}{s \tau_M}} 
\right), \label{eq:ebm} \\
J(t) &= J_0 \left({1 + \Delta \int_0^\infty \left(1 - \e^{-\gamma t}\right)
\rho(\gamma) \; \d \gamma + \frac{t}{\tau_M}} 
\right),
\end{align}
for a density function (relaxation spectrum) $\rho(\gamma)$ as a function of 
frequency, which  
satisfies $\rho(\gamma) \geq0$ and
\begin{equation}
 \int_0^ \infty \rho(\gamma) \; \d \gamma = 1.
\end{equation}
The Burgers model in \eqref{eq:burgers} is a special case of \eqref{eq:ebm} 
with a single frequency, $\rho(\gamma)= \delta(\gamma - 1/\tau_r)$, where 
$\delta(x)$ is the Dirac delta function. The integral in \eqref{eq:ebm} can be 
recognised as the Stieltjes transform of $\gamma \rho(\gamma)$. 

\begin{figure}
\centering
 \includegraphics[width=0.6\columnwidth]{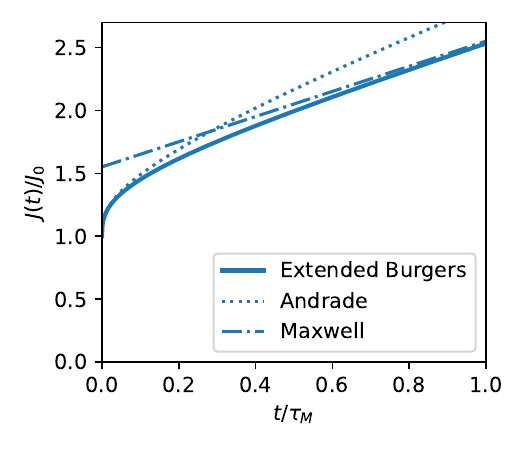}
 \caption{Scaled creep function $J(t)/J_0$ as a function of dimensionless time 
$t/\tau_M$ for an extended Burgers model with the parameters given in Table 1 
for tetrakaidecahedral grains ($\alpha=\tfrac{1}{3}$, $\tau_A/\tau_M = 1.68$, 
$\Delta = 0.552$). The short-time behaviour is well-described by an Andrade 
model and the long-time behaviour by a Maxwell model.}
 \label{fig:time_domain}
\end{figure}

The extended Burgers model in \eqref{eq:ebm} is sufficiently general to 
describe any linear viscoelastic response. A commonly-used parametrisation of
$\rho(\gamma)$ has a power-law distribution in a period band between a 
short-period cut-off $\tau_L$ and a long-period cut-off $\tau_H$ 
(e.g. \citep{Anderson1979,Jackson2010}). In the 
present work we will neglect the short-period cut-off, and write the density as
\begin{equation}
 \rho(\gamma) = \begin{cases}
                 \dfrac{\alpha \tau_H}{(\gamma \tau_H)^{\alpha + 1}}, 
\quad \quad &\gamma  \tau_H > 1,\\
                 0, \quad \quad 
&\text{otherwise.}
                \end{cases}
\end{equation}
in terms of the power law exponent $\alpha$. The Stieltjes transform can be 
determined analytically, and \eqref{eq:ebm} can be written as
\begin{equation}
 s \tilde{J}(s) = J_0 \left({1 + \Delta\, {}_2 F_1 \left(1, \alpha, 1+ \alpha, 
-s \tau_H \right) + \frac{1}{s \tau_M}} 
\right)
\end{equation}
where ${}_2 F_1$ is the ordinary hypergeometric function. The corresponding 
creep function (plotted in \autoref{fig:time_domain}) is
\begin{equation}
 J(t) = J_0 \left( 1 + \Delta \left(1 - \alpha\, E_{1+\alpha} 
\left(\frac{t}{\tau_H}\right)\right) + \frac{t}{\tau_M} \right)
\end{equation}
where $E_n(x)$ is the generalized exponential integral.

Note that as $s \tau_H \rightarrow \infty$ (high frequencies)
\begin{equation}
 {}_2 F_1 \left(1, \alpha, 1+ \alpha, -s 
\tau_H \right) \sim \frac{\Gamma(1+ \alpha) \Gamma(1- \alpha)}{(s 
\tau_H)^\alpha}
\end{equation}
and so the extended Burgers model behaves in the same way as the Andrade model 
at high frequencies, with parameters related by
\begin{equation}
 \tau_A = \tau_H \left( \Gamma(1-\alpha) \, \Delta  \right)^{-1/\alpha}.
\end{equation}
As $s \tau_H \rightarrow 0$ (low frequencies), 
\begin{equation}
 s \tilde{J}(s) \sim J_0 \left({1 + \Delta  + \frac{1}{s \tau_M}} 
\right)
\end{equation}
and the extended Burgers model behaves as a Maxwell material.

\section{Voigt averaging}\label{sec:voigt}

Figures \ref{fig:tetra_moduli} and \ref{fig:tetra_all} show results for 
tetrakaidecahedral grains with Voigt-averaged properties. The 
Voigt-averaged shear modulus is given in the Laplace domain by
\begin{equation}
 s \tilde{G}^{V} (s) = \frac{2}{5} s \tilde{G}^{(1)} (s) + \frac{3}{5} s 
\tilde{G}^{(2)} (s),
\end{equation}
where superscript $V$ denotes the Voigt-average, and superscripts (1) and (2) 
represent deformation with different principal axes: (1) has principal axes 
aligned with the square faces (see equation (40) of \citep{Rudge2018a}). It 
should be noted that the quantities presented in \autoref{fig:tetra_moduli} 
average in slightly different ways, although the averaging formulae are a 
direct consequence of the above expression. For example, for $s \ll 1$ the 
material behaves as a Maxwell model for which
\begin{equation}
 s \tilde{G}(s) \sim s \eta - \frac{s^2 \eta^2}{G_{ss}} + \cdots 
\end{equation}
and hence the steady-state viscosity is Voigt-averaged as
\begin{equation}
 \eta^V = \frac{2}{5} \eta^{(1)} + \frac{3}{5} \eta^{(2)}
\end{equation}
whereas the steady-state modulus is Voigt-averaged as
\begin{equation}
 \frac{(\eta^V)^2}{G_{ss}^V} = \frac{2}{5} \frac{(\eta^{(1)})^2}{G_{ss}^{(1)}}  
+ \frac{3}{5} \frac{(\eta^{(2)})^2}{G_{ss}^{(2)}}.
\end{equation}
Similarly, for $s \gg 1$ the material behaves as an Andrade model with
\begin{equation}
 s \tilde{G}(s) \sim G_0 - \frac{G_0 \Gamma(1+\alpha)}{\left(s 
\tau_A\right)^\alpha} + \cdots 
\end{equation}
and thus the unrelaxed modulus averages as
\begin{equation}
G_0^V = \frac{2}{5} G_0^{(1)} + \frac{3}{5} G_0^{(2)},
\end{equation}
while the Andrade time averages as
\begin{equation}
 \frac{G_0^V}{\left(\tau_A^V\right)^\alpha} = \frac{2}{5} 
\frac{G_0^{(1)}}{\left(\tau_A^{(1)}\right)^\alpha}  + \frac{3}{5}  
\frac{G_0^{(2)}}{\left(\tau_A^{(2)}\right)^\alpha}.
\end{equation}
It should also be noted that the Voigt-averaged Maxwell time is defined by 
$\tau_M^V = \eta^V / G_0^V$.

\section{Rational fits to functions of Poisson's ratio}\label{sec:rational}

The data shown in Figures 
\ref{fig:hex_moduli} and \ref{fig:tetra_moduli} as a function of Poisson's 
ratio $\nu$ can be effectively parametrised by fitting a rational function of 
the form
\begin{equation}
 r(\nu) = \frac{\displaystyle \sum_{j=1}^{m} \frac{w_j f_j}{\nu 
-\nu_j}}{\displaystyle  \sum_{j=1}^{m} 
\frac{w_j}{\nu -\nu_j}}
\end{equation}
where $\nu_j$ are the nodes, $f_j$ are function values, and $w_j$ 
are weights. The rational function has the property that $r(\nu_j) = f_j$. The 
above equation expresses the rational function in barycentric form, and an 
alternative and equivalent expression is
\begin{equation}
 r(\nu) = \frac{\displaystyle \sum_{j=1}^{m} w_j f_j  \prod_{k \neq j} 
(\nu - \nu_k)}  {\displaystyle \sum_{j=1}^{m} 
w_j \prod_{k \neq j} 
(\nu - \nu_k) },
\end{equation}
from which it is clear the function is the ratio of two polynomials of degree 
$m-1$.  The 
weights can all be scaled by a constant factor and the rational function will 
be unchanged. To specify the weights uniquely, here we fix $w_1 = 1$.

\begin{table}[h!]
\centering
 \begin{tabular}{c|cc|c}
 & $f_1$ & $f_2$ & $w_2$\\ \hline 
 $G_0/\mu$ & 0.599973 & 0.857133 & -0.700034\\
 $G_{ss}/\mu$ & 0.353374 & 0.686123 &  -0.515031\\
 $(\tau_A/\tau_M)^\alpha$ & 0.68953 & 1.319737 & -0.301846
 \end{tabular}
\caption{Parameters of a rational fit to the data shown in 
\autoref{fig:hex_moduli} as a function of $\nu$ for hexagonal 
grains. In each case $\nu_1 = -1$, $\nu_2 = \tfrac{1}{2}$, and $w_1=1$. The 
Andrade exponent $\alpha = 
0.3672092$.}
\label{tab:hex_rational}
\end{table}

\autoref{tab:hex_rational} gives parameters which fit the data for 
hexagons shown in \autoref{fig:hex_moduli}. The fits have just three free 
parameters, and represent the properties as a ratio of two linear functions of 
$\nu$ (i.e. $m=2$). The 
fits for the scaled moduli $G_0/\mu$ and $G_{ss}/\mu$ have relative errors 
$<10^{-6}$ which suggests that the representation of these quantities in terms 
of a ratio of linear functions may in fact be exact, similar to the analytic 
result for $G_0/\mu$ for spheres by Zener \citep{Zener1941}. Larger errors are 
seen 
in fitting the scaled Andrade time $\tau_A/\tau_M$, but the parameters in the 
table still represent an excellent fit with relative errors $<10^{-3}$.

\begin{table}[h!]
\centering
 \begin{tabular}{c|ccc|cc}
 & $f_1$ & $f_2$ & $f_3$ & $w_2$ & $w_3$  \\ \hline 
 $G^{(1)}_{0}/\mu$ & 0.249839 & 0.74434 & 0.556733 & 1.404701 & -2.491113\\
 $G^{(2)}_{0}/\mu$ & 0.611883 & 0.873231 & | & -0.791083 & | \\
 $G^{V}_{0}/\mu$ & 0.46641 & 0.821582 & 0.685598 & 0.97236 & -1.981183 \\
 $G^{(1)}_{ss}/\mu$ & 0.029394 & 0.498677 & 0.265034 & 1.428164 & -2.599101 
  \\
 $G^{(2)}_{ss}/\mu$ & 0.135534 & 0.693088 & 0.491729 & 2.490243 & -3.739137 
\\
 $G^{V}_{ss}/\mu$ &  0.089789 & 0.56882  & 0.390075 & 2.503946 & -3.76036
\\
 $(\tau^{(1)}_A/\tau_M^{(1)})^\alpha$ & 0.188561 & 1.176442 & 0.696462 & 
22.349757 & -32.371228 \\
 $(\tau^{(2)}_A/\tau_M^{(2)})^\alpha$ & 0.35751 & 1.862253 & 1.123438 & 
1.583196 & -2.715582 \\
 $(\tau^{V}_A/\tau_M^{V})^\alpha$ & 0.308609 & 1.444995 & 0.901722 & 10.238062 &
-14.343569\\
 \end{tabular}
\caption{Parameters of a rational fit to the data shown in 
\autoref{fig:tetra_moduli} as a function of $\nu$ for 
tetrakaidecahedral 
grains. In each case $\nu_1 = -1$, $\nu_2 = \tfrac{1}{2}$, $\nu_3=0$, and 
$w_1=1$. The 
Andrade exponent $\alpha = 
\tfrac{1}{3}$.}
\label{tab:tetra_rational}
\end{table}

\autoref{tab:tetra_rational} gives parameters for fits of the data plotted in 
\autoref{fig:tetra_moduli} for tetrakaidecahedral grains. Unlike the case of 
hexagons, the data require higher order polynomials to be fit well; the table 
uses a fit of a ratio of quadratics ($m=3$), which fits the scaled moduli to 
within a 
relative error of $<10^{-3}$, and the scaled Andrade times $\tau_A/\tau_M$ to 
within $<2 \times 10^{-2}$.

\section{Scaling argument for high frequency attenuation}\label{sec:lee_scaling}

The scaling argument given here follows that in Lee et al. \citep{Lee2011}. The 
balance of terms in \eqref{eq:bdry_diff3} determines a natural lengthscale $l$ 
over which grain boundary diffusion acts when forced at a frequency $\omega$. 
For a typical strain $\epsilon$, the surface tractions are of magnitude 
$p \sim \mu \epsilon$, and jumps in the normal velocity of magnitude $\left[ 
\dot{\u} \cdot \n \right] \sim \epsilon l \omega$. Substituting these scalings 
into $\eqref{eq:bdry_diff3}$, with the surface Laplacian scaling as $1/l^2$, 
yields the expression for $l$ given in \eqref{eq:ldef}.

The energy stored 
elastically per unit volume scales as
\begin{equation}
 \mathcal{U} \sim \mu \epsilon^2.
\end{equation}

The energy dissipated by grain boundary diffusion over a cycle scales as
\begin{equation}
 \mathcal{D} \sim \frac{\Omega \delta 
D^\text{gb}}{k T \omega V} \int_{S_i} |\nablas p|^2 
\; \d S = \frac{l^3}{\mu V} \int_{S_i} |\nablas p|^2 
\; \d S. \label{eq:energy_diss}
\end{equation}

With a stress singularity at the triple junction, the normal stress is
\begin{equation}
 p \sim \mu \epsilon \left(r /d \right)^{-\lambda}
\end{equation}
where $d$ is a measure of grain size, and $r$ is distance from the triple 
junction. Thus the gradient scales as
\begin{equation}
\nablas p \sim \frac{\mu \epsilon}{d} \left(r /d \right)^{-(\lambda + 1)}.
\end{equation}

The surface integral term in \eqref{eq:energy_diss} can be approximated in 2D 
as
\begin{equation}
 \int_l^\infty |\nablas p|^2 \, \d r \sim \frac{\mu^2 \epsilon^2}{d} \left(l /d 
\right)^{-(2\lambda + 1)}
\end{equation}
where it is assumed that diffusion smooths the singularity within a distance 
$l$ of the triple junction, and that $l \ll d$. In 2D the grain volume $V \sim 
d^2$ and hence \eqref{eq:energy_diss} becomes
\begin{equation}
 \mathcal{D} \sim \mu \epsilon^2 \left(l/d \right)^{2(1 - \lambda)} \sim \mu 
\epsilon^2 \left(\omega \tau_M \right)^{-2(1 - \lambda)/3}. 
\label{eq:diss_express}
\end{equation}

The attenuation can be related to the ratio of the energy dissipated over a 
cycle to the energy stored and hence
\begin{equation}
 Q^{-1} \sim \frac{\mathcal{D}}{\mathcal{U}} \sim \left(\omega \tau_M 
\right)^{-\alpha}
\end{equation}
where
\begin{equation}
 \alpha = \frac{2}{3} \left(1- \lambda \right). \label{eq:lam_alpha}
\end{equation}

\section{Volume diffusion}\label{sec:vdiff}

The main focus of this work has been on the process of grain boundary 
diffusion, as that process
has been suggested to be the one of relevance in a number of  
experimental studies \citep[e.g.][]{Takei2017}. However, there are 
circumstances where diffusion within the interiors of grains dominates over 
diffusion on the grain boundaries, and in steady-state this is known 
as 
Nabarro-Herring creep. From a modelling perspective volume diffusion can be 
treated 
in a very similar way to surface diffusion, and simply results in replacing a 
surface integral in the 
weak form of the equations by a volume integral, i.e. instead of 
\eqref{eq:weak_diff3}, one has
\begin{equation}
-\int_{S_\text{grain}} \hat{\u}^* \cdot 
\n \, q
\;\d S + \frac{\Omega D}{k T s^*} \int_{V_\text{grain}} 
\nabla p^* \cdot 
\nabla q 
\; \d V = \frac{1}{2} \int_{S_\text{grain}} \left(\n \cdot \tensor{\Gamma}^* 
\cdot \R \right) q
\;\d S 
\label{eq:weak_diff_vol}
\end{equation}
where $D$ is the self-diffusion coefficient for volume diffusion (see 
\citep{Rudge2018a,Rudge2021}). 

\begin{figure}
 \includegraphics[width=\columnwidth]{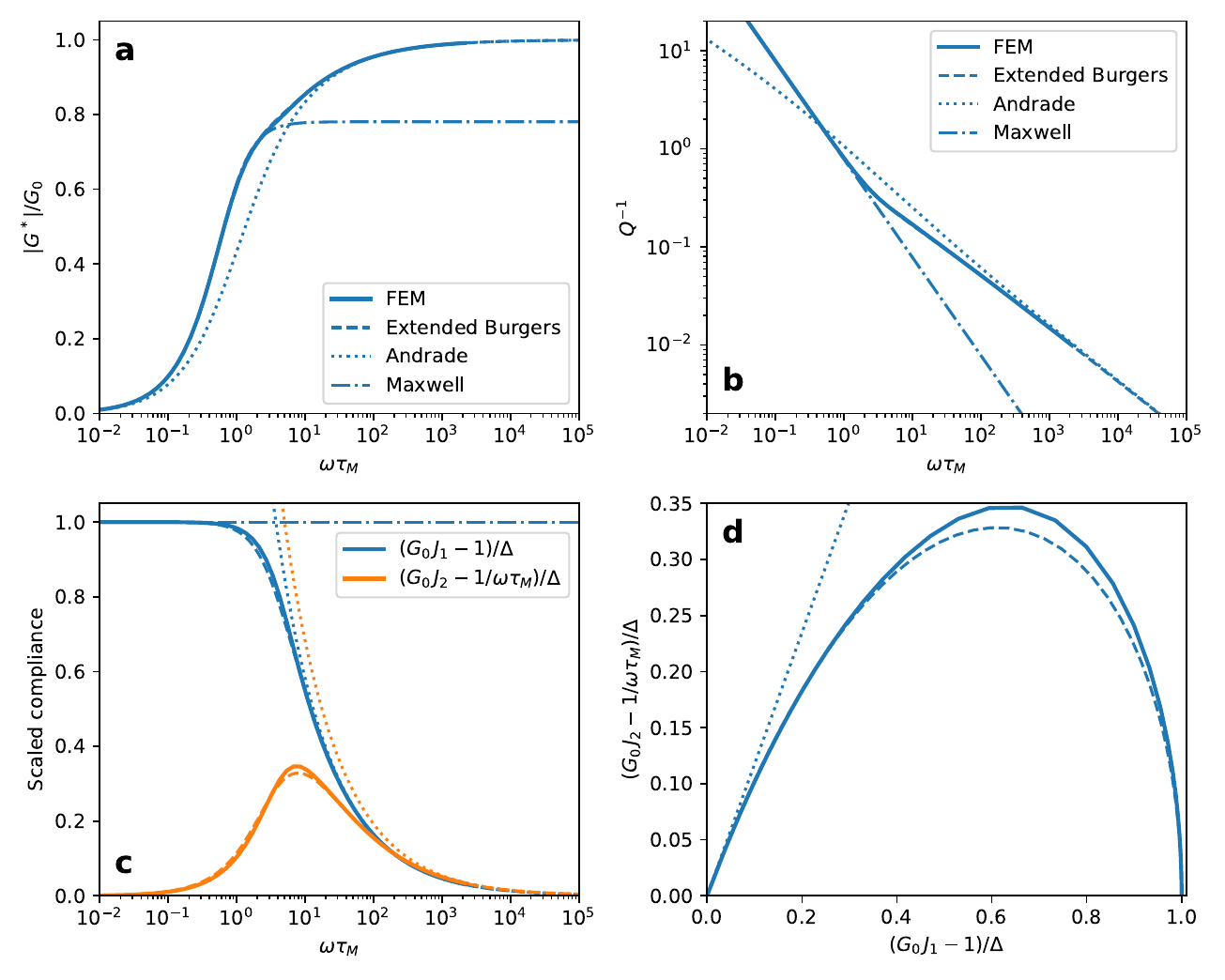}
 \caption{Plots similar to \autoref{fig:hex_all} for hexagonal grains with 
volume diffusion rather than surface diffusion. The extended Burgers fit has 
$\Delta=0.281$, $\alpha=0.5508138$ and $\tau_A/\tau_M = 0.99$.}
 \label{fig:hex_volume}
\end{figure}

\autoref{fig:hex_volume} illustrates the viscoelastic behaviour associated with 
volume 
diffusion for the case of hexagons. The results are again scaled using a 
Maxwell time $\tau_M = \eta / G_0$ where $G_0$ is exactly the same as before 
(as the pure elastic problem is unchanged) but the steady-state viscosity is 
different and given by
\begin{equation}
\eta = \frac{5}{288} \frac{k T d^2}{D \Omega},
\end{equation}
proportional to the grain size squared rather than cubed. The above 
expression has a different numerical prefactor to the expression for 
hexagons given in equation (36) of \citep{Rudge2018a} (a prefactor of 1/36). 
This prefactor difference is a consequence of the 
difference in homogenisation scheme, with the scheme used here being the same 
as 
the equivalent dissipation principle in \citep{Rudge2021}. 

The principal difference in viscoelastic behaviour between grain boundary 
diffusion and volume diffusion concerns the Andrade exponent $\alpha$, which 
can be 
understood using a similar scaling argument to that presented in 
\autoref{sec:lee_scaling}. The length scale over which volume diffusion acts 
when 
forced at a frequency $\omega$ is
\begin{equation}
 l = \left(\frac{\mu D \Omega}{k T \omega} \right)^{1/2},
\end{equation}
which replaces the surface diffusion result in \eqref{eq:ldef}. The expression 
for dissipation in 
\eqref{eq:energy_diss} becomes
\begin{equation}
 \mathcal{D} \sim \frac{\Omega D}{k T \omega V} \int_{V} |\nabla p|^2 
\; \d V = \frac{l^2}{\mu V} \int_{V} |\nabla p|^2 
\; \d V \label{eq:energy_diss_vol}
\end{equation}
and the expression in \eqref{eq:diss_express} becomes
\begin{equation}
 \mathcal{D} \sim \mu \epsilon^2 \left(l/d \right)^{2(1 - \lambda)} \sim \mu 
\epsilon^2 \left(\omega \tau_M \right)^{-(1 - \lambda)},
\end{equation}
and hence the expression relating the Andrade exponent to 
the stress exponent in \eqref{eq:lam_alpha} becomes
\begin{equation}
 \alpha =  1- \lambda. \label{eq:lam_alpha_vol}
\end{equation}
Thus on attenuation plots such as \autoref{fig:hex_volume}b volume 
diffusion shows a steeper slope than grain boundary diffusion does at high 
frequencies (steeper by a factor of 3/2).

\section*{Acknowledgments}

I thank H. Yamauchi and Y. Takei for kindly sharing their data. I thank two 
anonymous reviewers for their helpful comments. I also thank H. 
Innes, G. Johnson, T. Breithaupt, 
D. Al-Attar, Y. Takei and D. McKenzie for insightful discussions on transient 
creep.

\section*{Data accessibility}

Python code used for the finite element simulations is available at 
\url{https://github.com/johnrudge/transient_diffusion_creep}. The results of 
the finite element calculations (used in plotting the figures) can be found 
in a spreadsheet 
in the supplemental information.

\section*{Funding statement}

This research received no specific grant from any funding agency in the public, 
commercial, or not-for-profit sectors.


\bibliographystyle{RS}

\end{document}